%% file: QPO_FTP.tex
\documentclass[article]{aa}

\usepackage{graphicx}
\usepackage{txfonts}

\usepackage[colorlinks=true, citecolor=blue, linkcolor=blue, urlcolor=blue]{hyperref}

\begin{document}

    \title{Morphology-aware periodicity searches in \textit{Fermi}-LAT blazars}

    \titlerunning{Morphology-aware periodicity searches in \textit{Fermi}-LAT blazars}

\author{A. A. Maldonado,\inst{1}\fnmsep\thanks{aarona01@ucm.es}
        A. Dom\'inguez,\inst{1,2}\fnmsep\thanks{alberto.d@ucm.es}
        A. Dinesh,\inst{1,2}
        A. Rico,\inst{3}
        P. Pe\~nil,\inst{3}
        S. Buson,\inst{4,5}
        \and M. Ajello\inst{3}
        }

\authorrunning{A. A. Maldonado et al.}

\institute{$^1$Department of EMFTEL, Universidad Complutense de Madrid, E-28040 Madrid, Spain \\
            $^2$Instituto de Física de Partículas y del Cosmos (IPARCOS), Universidad Complutense de Madrid, E-28040 Madrid, Spain \\
            $^3$Department of Physics and Astronomy, Clemson University, Kinard Lab of Physics, Clemson, SC 29634-0978, USA \\
            $^4$Deutsches Elektronen-Synchrotron DESY, Platanenallee 6, 15738 Zeuthen, Germany \\
            $^5$Julius-Maximilians-Universität Würzburg, Fakultät für Physik und Astronomie, Emil-Fischer-Str. 31, D-97074 Würzburg, Germany}

   \date{Received April 2, 2026}

\abstract
{Standard periodicity searches in active galactic nuclei assume sinusoidal modulation, whereas $\gamma$-ray blazar emission is often dominated by asymmetric flares and localized bursts that a harmonic basis matches poorly.}
{We quantify how the significance of a periodic signal recovered from \textit{Fermi}-LAT light curves depends on the assumed intra-cycle morphology and on the data-quality cuts applied to the light curves.}
{We analyze 101 seven-day binned light curves from the \textit{Fermi}-LAT Light Curve Repository (LCR) with the Fast Template Periodogram, using sinusoidal, narrow Gaussian burst, and empirical asymmetric templates. Significances are calibrated with red-noise simulations that include a Poisson noise floor. A Random Forest classifier trained on periodogram, phase-coherence, and period-stability features scores the sources.}
{A few bins with unconstrained likelihood fits flatten the periodogram and make the simulated null effectively white. With the quality cuts, three sources exceed a per-source significance of $3\sigma$, against $0.14$ expected by chance. One of them is the established candidate PG 1553+113, and none reaches $3\sigma$ after the trial correction for the 101 light curves searched. The template changes the recovered significance substantially: for 4FGL J0102.8+5824 it is $0.24\sigma$ with a sinusoid and $3.85\sigma$ with a Gaussian burst. A gain of this kind is expected for about one source in simulated noise, and about $40\%$ of the sources are best described by the sinusoid. The significances have an upper tail heavier than the Gaussian null predicts, consistent with flares.}
{We report no new periodicity detection. Non-sinusoidal templates can change the recovered significance substantially and usefully complement the sinusoid, but a large gain in significance does not by itself establish a periodic signal. The classifier score responds to isolated flares as well as to periodic signals, another sign of flares absent from the null. Our results agree with recent systematic searches. Applying the convergence criterion of the LCR documentation, complemented by a simple consistency check, is important for periodicity searches, especially with 7-day bins.}


   \maketitle
\nolinenumbers

\section{Introduction}

The continuous monitoring of the $\gamma$-ray sky by the Large Area Telescope (LAT) on board the \textit{Fermi} Gamma-ray Space Telescope \citep[\textit{Fermi}-LAT;][]{atwood2009} has revolutionized the study of active galactic nuclei (AGN) in the time domain. Over the past decade, several year-scale quasi-periodic oscillation (QPO) candidates have been identified in blazar light curves, with PG 1553+113 \citep[e.g.,][]{ackermann2015, tavani2018, abdollahi2024, penil2024} and PKS~2155$-$304 \citep[e.g.,][]{sandrinelli2014,penil2020} serving as benchmark cases. The physical mechanisms driving these modulations remain under debate, but proposed models broadly fall into two major scenarios. The first involves macroscopic geometric modulations, where the viewing angle and relativistic Doppler factor change periodically, such as the orbital motion in supermassive black hole binaries \citep[SMBBHs; e.g.,][]{sillanpaa1988, begelman1980}, Lense-Thirring precession \citep[e.g.,][]{caproni2013}, or helical jet kinematics \citep[e.g.,][]{camenzind1992}. The second involves plasma-driven scenarios, where the actual particle acceleration or magnetic dissipation is periodically pulsed (e.g., periodic magnetic reconnection within the jet flow; \citealt{giannios2009}).

To date, the search for blazar QPOs has been overwhelmingly dominated by standard frequency-domain algorithms, most notably the Lomb-Scargle Periodogram \citep[LSP;][]{lomb1976,scargle1982}. While computationally efficient and well-suited for unevenly sampled data, the LSP implicitly assumes that the underlying periodic signal is sinusoidal. Mathematically, it projects the data onto a basis of sine and cosine functions. However, blazar $\gamma$-ray emission is often non-linear and can be characterized by sharp asymmetric flares, fast-rise exponential-decay (FRED) profiles, or localized bursts separated by long quiescent periods.

Time-frequency analyses, such as wavelet transforms, characterize the lifetime and frequency drift of a QPO, but they are typically applied at timescales where the intra-cycle details of the periodic signal are averaged out. Sinusoidal methods suffer from spectral leakage for the burst-like flares typical of blazars: the power of a non-sinusoidal signal is spread over several frequencies, which dilutes the significance of the fundamental period \citep[e.g.,][]{vaughan2005}.

To overcome this limitation, periodicity searches should be sensitive to a broader range of light-curve shapes. The Fast Template Periodogram \citep[FTP;][]{hoffman2017} generalizes the Lomb-Scargle approach to accommodate arbitrary non-sinusoidal periodic templates. By cross-correlating the time series with a user-defined template, the FTP can increase the response to signals whose morphology is not well described by a sine wave.

In this work, we investigate how the significance of $\gamma$-ray blazar periodicities depends on the assumed intra-cycle light-curve profile. We evaluate 101 seven-day binned light curves from the \textit{Fermi}-LAT Light Curve Repository \citep[LCR;][]{abdollahi2023} against three templates: a sinusoid, a narrow Gaussian burst, and an empirical asymmetric template derived from PG~1553+113. Because the calibration of any periodicity search depends on the quality of its input light curves, we also quantify the effect of data-quality cuts on the LCR light curves.

A high peak in a periodogram can also arise from stochastic red-noise flares that happen to align in phase \citep[e.g.,][]{vaughan2016, covino2019, covino2020}. The false-alarm probability (FAP) measures how often noise produces a peak of a given height, but not whether the signal persists across the observing baseline. We therefore complement the FTP with a Random Forest machine learning (ML) classifier, trained on simulated light curves, that combines periodogram features with measures of phase coherence and period stability. We use it to test whether the sources with high significances also show phase coherence, not to confirm or reject individual signals, and we test what its score responds to.

The paper is structured as follows. Section~\ref{sec:data} describes the sample, the quality cuts, the templates, the significance calibration, and the classifier. Section~\ref{sec:results} presents the effect of the quality cuts, the number of sources crossing the significance threshold, the dependence of the significance on the template, and the classifier ranking. Section~\ref{sec:discussion} discusses these results in the context of previous work, Sect.~\ref{sec:physics} outlines physical scenarios that could produce burst-dominated periodic profiles, and Sect.~\ref{sec:conclusions} summarizes our conclusions.

\section{Data and methods}
\label{sec:data}

\subsection{Sample and data quality}
\label{subsec:sample}

The LCR provides light curves built from the data of the LAT on board the \textit{Fermi} Gamma-ray Space Telescope for the sources of the fourth \textit{Fermi}-LAT source catalog, data release 2 \citep[4FGL-DR2;][]{abdollahi2020, ballet2020}. We used the 7-day binned light curves (0.1--100 GeV), computed with a fixed photon index, between Modified Julian Dates 54686.5 and 61105.5, a baseline of 17.6 years, and identify sources by their 4FGL designations. For each bin, the LCR provides the flux, its uncertainty, the Test Statistic (TS) of the source, and a flag indicating whether the likelihood fit converged. \citet{abdollahi2023} caution that results from fits that did not converge should be considered suspect and should not be used for higher-level analyses, including power spectral density (PSD) and variability studies.

We applied three data-quality cuts. First, we rejected bins whose likelihood fit did not converge. Second, we rejected bins with $\text{TS} \geq 2$ whose flux uncertainty is zero or not finite. Third, we applied an internal-consistency test to the remaining bins. At fixed exposure and cadence, $F/\sqrt{\text{TS}}$ traces the sensitivity of each bin and is close to constant across a light curve. A bin with a large flux and a small TS therefore has an unconstrained likelihood. We rejected bins in which this ratio exceeds ten times its median over the light curve. A flare raises $F$ and TS together and leaves the ratio largely unchanged, which makes the test insensitive to genuine flaring. Across the sample, the median flux of the bins rejected by this test is 1.5 times its own uncertainty, and only seven of them have a flux more than three times their uncertainty. Varying the threshold between five and twenty times the median changes the number of rejected bins from 138 to 22, but the bins with the most extreme fluxes are removed in all cases.

Bins with $\text{TS} < 2$ were treated as non-detections. They were excluded from the periodograms and from the classifier features, and were used only through interpolation when estimating the long-term trend described below. We retained the light curves in which non-detections make up less than 30\% of the bins after the cuts, which yielded 101 sources. This limit also keeps the fraction of missing data well below the level at which spurious signals near one year, produced by the window function of the \textit{Fermi}-LAT observations, become important \citep{penil2025b}.

Prior to the periodicity analysis, we isolated the oscillatory component by removing the secular baseline drift via Singular Spectrum Analysis \citep[SSA;][]{broomhead1986, vautard1989}. We adopted a window length of $L = 0.4N$. To ensure objectivity, the secular trend was identified dynamically by grouping all leading reconstructed components whose zero-crossing timescales exceeded 6 years. To prevent artificial red-noise leakage at this detrending boundary, we restricted our subsequent periodicity search grid to a maximum period of 5.5 years. Subtracting this combined low-frequency component removes the baseline drift. All 101 light curves yielded stable baseline extractions, which we verified by varying $L$ between $0.3N$ and $0.5N$. SSA has also been found to be more robust to flares than the Lomb-Scargle periodogram, the continuous wavelet transform, and phase dispersion minimization \citep{penil2025d}.

\subsection{The FTP and the templates}
\label{subsec:templates}

Unlike the LSP, the FTP analytically computes the best fit of an arbitrary periodic template $T(\phi)$. The algorithm optimizes the amplitude, frequency, phase shift $\phi_0$, and constant offset $C$. Because the offset and amplitude are free parameters, the periodogram power does not depend on the zero-mean and unit-amplitude normalization of the templates. The free phase shift $\phi_0$ makes the result independent of their absolute phase alignment. We searched 500 trial periods between 1 and 5.5 years.

We evaluated the light curves against three distinct template families, chosen to represent different phenomenological regimes. (1) A Sinusoidal Template, $T(\phi) = \sin(2\pi\phi)$, serves as the standard baseline, representing smooth, continuous geometric variations such as idealized Doppler boosting from a precessing jet. (2) A Gaussian Template represents a narrow, symmetric burst profile. We adopted a fixed fractional width of $\sigma = 0.05$ to specifically target highly localized flares or collimated sweeping effects that are poorly described by a broad sinusoid. A Gaussian with $\sigma = 0.05$ has a full width at half maximum (FWHM) of $\approx 12\%$ of the phase cycle. For a typical year-scale blazar periodicity (e.g., $P \sim 3$ years), this corresponds to an active emission duty cycle of a few months, physically consistent with the macroscopic $\gamma$-ray flaring timescales resolved by our 7-day binning. We deliberately excluded broader Gaussian templates (e.g., $\sigma > 0.05$). Because their FWHM spans $\gtrsim 35\%$ of the phase cycle, broader Gaussians morphologically degenerate into continuous harmonic waves, rendering them largely redundant with our standard sinusoidal baseline. Restricting the burst template to a single narrow profile keeps the search sensitive to discrete events while limiting the number of trials. (3) An Empirical Archetype serves as a purely observational template derived from the benchmark $\gamma$-ray QPO source PG 1553+113.

\subsection{Epoch-folding the archetype}
Blazar flares are commonly described by exponential profiles with different rise and decay times \citep[e.g.,][]{cerruti2025}, and can show complex secondary peaks \citep[e.g.,][]{penil2026b}. To search for sources with light-curve profiles similar to that of PG 1553+113, we generated an empirical template from the source itself.

We applied an epoch-folding technique to the 17.6-year, SSA-detrended light curve of PG~1553+113. Using its established period of $P \approx 2.1$ years \citep{ackermann2015,abdollahi2024}, we computed the phase $\phi = (t \bmod P) / P$, which folds the $\sim 8$ cycles of the source onto $\phi \in [0,1]$. We then grouped the phase-folded data into 15 uniform bins, with an average of $\sim 60$ observations per bin, and computed the mean flux within each bin.

This bin-averaging process suppresses part of the stochastic cycle-to-cycle fluctuations and provides an empirical estimate of the average intra-cycle modulation. The resulting folded profile was smoothed via cubic spline interpolation and normalized to zero mean and unit amplitude. Visual inspection confirmed this did not introduce artificial ringing. This empirical archetype, characterized by an asymmetric primary peak, was then used as one of the templates in the FTP analysis of the rest of the blazar sample.

We acknowledge the inherent circularity of evaluating PG 1553+113 against its own empirical profile. Its inclusion in the sample serves as a self-consistency check of the algorithm. For the remaining 100 sources, the template functions purely as an observational archetype for FRED-like asymmetry. Because this profile is derived exclusively from the independent flux history of PG 1553+113, it carries no a priori information regarding the noise realizations, observational gaps, or intrinsic kinematics of the other blazars. Applying this archetype to the rest of the sample therefore introduces no statistical contamination: it acts as a fixed basis function, in the same way as the Gaussian and the sinusoid.

\subsection{Significance calibration}
\label{subsec:fap}
We calibrated the FAP of each template with Monte Carlo simulations of red noise. For each source, we performed the following steps.

\begin{enumerate}
    \item We fitted the periodogram of the light curve, binned logarithmically into 25 frequency bins up to $1/21$~d$^{-1}$, with
\begin{equation}
P(\nu) = A\,\nu^{-\alpha} + C ,
\label{eq:psd}
\end{equation}
a power law together with a constant term representing the Poisson noise floor \citep{penil2025a, rico2025}, with $\alpha$ restricted to $[0,3]$. We refer to the ratio between the binned powers at the lowest and highest frequencies of the fit as the dynamic range of the periodogram. A value near unity indicates a flat periodogram, for which $\alpha$ is not constrained by the data. After the quality cuts, the median slope of the 101 light curves is $0.95$, with 68\% of the sources between $0.59$ and $1.31$. Four sources have $\alpha > 2.5$, three of them at the bound. Their periodograms are flat except for the two to four lowest-frequency bins, which carry long-term trends on timescales of 3 to 9 yr, and the steep slope describes those bins. Refitting their PSDs only above $1/(6~\text{yr})$, the frequencies that survive the detrending, and recomputing the significances with 50 to 70 realizations, changes the per-source significance by less than $0.2\sigma$ for three of them and lowers it for the fourth, and all four remain below zero. A simple power law provides a uniform, minimally parameterized baseline against which all templates are tested. It is not the only plausible null for burst-dominated light curves. Light curves generated with the \citet{timmer1995} algorithm contain flare-like excursions on all sampled timescales, but they are Gaussian and do not reproduce individual FRED flares.
    \item Using the \citet{timmer1995} algorithm, we generated synthetic light curves from the fitted PSD, including the constant term, and rescaled them to the observed flux variance. Each synthetic light curve was sampled at the observed epochs, masked with the gaps and non-detections of the observed light curve, and detrended with the same SSA procedure, including the interpolation over non-detections, before the FTP evaluation.
    \item For each template, the single-template FAP, $\text{FAP}_{1}$, is the fraction of synthetic light curves whose maximum FTP power in the searched band equals or exceeds the observed maximum. Because we test three templates on each light curve, we applied a \v{S}id\'{a}k correction, $\text{FAP}_{\rm src} = 1 - (1 - \text{FAP}_{1})^3$, and converted $\text{FAP}_{\rm src}$ into a one-sided Gaussian-equivalent significance, $\sigma_{\rm src}$, which we call the per-source significance. It is negative when $\text{FAP}_{\rm src} > 0.5$, and we keep the sign because $\Delta\sigma$ (Sect.~\ref{subsec:dsigma}) is a difference of such values. Global significances below $1\sigma$ carry no evidential weight, and their exact values are set by the steep mapping of FAPs close to unity onto $\sigma$; we report them as $<1$, as in \citet{rico2025}. It is the significance of the best peak found in a single light curve, after the search over the periodogram band, which is built into the maximum-power statistic, and over the three templates. The three template families were fixed a priori on phenomenological grounds, and the correction over three trials accounts for the complete template search space.
\end{enumerate}

Following \citet{penil2025c}, we call global significance the value that also accounts for the number of sources searched \citep[e.g.,][]{penil2025a}. We obtain it with the same correction over the 101 light curves, denote it $\sigma_{\rm glob}$, and report it alongside the per-source value. The per-source significance already includes the search over periods and is therefore not the pre-trial local significance of \citet{penil2025c}, which refers to a given period. The trial factor over periods enters the global significance in both cases.

We also report a local significance, $\sigma_{\rm loc}$, which refers to a given period. We include it because other studies of the same sources quote local significances \citep{rico2025, penil2026}, and it places our results on a footing closer to theirs. For each template, we compare the observed power at the period of the observed peak with the powers of the synthetic light curves at that same period, and convert the fraction of synthetic light curves at or above the observed power into a Gaussian-equivalent significance. No trial factor over periods or templates is applied, and we quote the highest of the three values. The period is that of the observed peak, and the local significance is not equivalent to the local values quoted in other works, which are obtained with different methods, binning, and baselines. Its ceilings are $3.29\sigma$ for $N = 2\times10^{3}$ and $4.26\sigma$ for $N = 10^{5}$, and values at a ceiling are lower limits.

With $N$ synthetic light curves and no simulated peak exceeding the observed one, the single-template FAP is limited to $1/N$ and the significance saturates. After the correction for three templates, the ceiling is $2.97\sigma$ for $N = 2\times10^{3}$ and $4.01\sigma$ for $N = 10^{5}$. These ceilings, and the $3\sigma$ threshold used below, refer to the per-source significance, which already includes the trial factors over periods and over the three templates. They are not equivalent to a local significance of the same nominal value quoted in other works, which does not include these trials: a per-source significance of $3\sigma$ is a more stringent condition than a pre-trial local significance of $3\sigma$. We therefore ran $2\times10^{3}$ realizations for every source in a first pass and repeated the analysis with $10^{5}$ realizations for the sources whose first-pass significance lay within $0.3\sigma$ of the ceiling, to allow for the Poisson uncertainty of the first-pass FAP.

Under a correctly calibrated null, a per-source threshold of $3\sigma$ corresponds to $\text{FAP}_{\rm src} = 1.35\times10^{-3}$, and $0.14$ sources are expected to exceed it in a sample of 101. Because the three templates are correlated, the correction over templates is conservative, and expected numbers derived from it are upper limits. We use the comparison between the observed and expected numbers as a check of the calibration. For the local significance, the expected number follows from applying the same rule to every synthetic light curve of the first pass, each compared with the other realizations of the same source, and summing over the 101 sources the fraction that reaches a given value.

\subsection{Structural ranking with an ML classifier}
\label{subsec:ml_setup}

We trained a supervised classifier on six structural features, computed over the same period range as the FTP search. The six features capture three distinct domains of the signal:
\begin{itemize}
    \item Temporal Stability: (1) \texttt{r\_period\_splits}: The ratio of the best-fit FTP period in the first half of the light curve versus the second half, penalizing transient or drifting signals.
    \item Frequency-Domain Topology: (2) \texttt{FTP\_prominence\_ratio}: The height of the primary periodogram peak relative to the local noise continuum. (3) \texttt{Power\_Ratio\_FTP\_vs\_LSP}: The ratio of the peak power recovered by the morphology-aware FTP versus the LSP.
    \item Signal-to-Noise \& Autocorrelation: (4) \texttt{SNR\_osc}: The amplitude of the oscillatory component relative to the residual noise. (5) \texttt{autocorrelation\_osc} and (6) \texttt{autocorrelation\_raw}: The first-lag autocorrelation coefficients of the isolated oscillatory component and the raw light curve, respectively, measuring cycle-to-cycle internal memory.
\end{itemize}
We did not include features computed from the binned phase-folded profile, because they depend on the choice of phase binning.

The training set consists of $2\times10^{4}$ simulated light curves built in flux units. Each takes the gap structure, flux uncertainties, PSD parameters, and flux standard deviation of one randomly chosen source of the sample, which places the features of the simulated and observed light curves on the same scale. Half of them are pure red noise generated from those PSD parameters. The other half contain in addition a strictly periodic signal with one of the three template shapes, a period between 1.2 and 5 yr, and an amplitude between 0.5 and 3 times the flux standard deviation.

A potential risk is data leakage. Because the positive class is built from the template shapes that the FTP searches for, a classifier that relies only on FTP power could memorize the periodogram response to its own templates instead of learning phase coherence. To mitigate this risk, the feature space combines two classes of metrics. We label as ``in-band'' the two features that are computed directly from the amplitude response of the periodograms and are therefore coupled to the template-fitting machinery: \texttt{FTP\_prominence\_ratio} and \texttt{Power\_Ratio\_FTP\_vs\_LSP}. We label as ``out-of-band'' the four features that do not depend on the amplitude response of any template fit: \texttt{r\_period\_splits}, which uses only the location of the periodogram maximum in each half of the light curve, and the time-domain statistics \texttt{SNR\_osc}, \texttt{autocorrelation\_osc}, and \texttt{autocorrelation\_raw}.

This training strategy has two limitations. First, the injected signals are strictly periodic, whereas astrophysical QPOs show cycle-to-cycle jitter in amplitude and phase. Second, the noise class is drawn from the same Gaussian red-noise model as the significance calibration and does not contain discrete flares. Section~\ref{subsec:ml_results} tests the response of the classifier to both.

We trained a Random Forest classifier \citep{breiman2001} on this balanced dataset. To prevent memorization and overfitting, the model was defined as an ensemble of 100 decision trees (\texttt{n\_estimators=100}), with depth and leaf-level regularization (\texttt{max\_depth=15}, \texttt{min\_samples\_leaf=5}) and balanced class weighting. The classifier returns a structural score, $S_{\rm morph}\in[0,1]$. This score is a ranking statistic learned on a synthetic training distribution and not a calibrated probability that a source hosts a periodicity. We use it to order the sources and not to accept or reject any of them.


\section{Results}
\label{sec:results}

\subsection{Effect of the quality cuts}
\label{subsec:qc}

Across the sample, the three cuts reject 135 bins with $\text{TS} \geq 2$: 74 with non-convergent fits, 17 more with a zero or non-finite flux uncertainty, and 44 identified by the consistency test. The last group amounts to 0.06\% of all bins with $\text{TS} \geq 2$. Some of the rejected bins have fluxes up to $\sim10^{6}$ times the median of their light curve.

An isolated spike has a flat power spectrum, and a few such bins are enough to flatten the periodogram of an entire light curve. A pure power law fitted to a flat periodogram returns an index consistent with zero, and with the constant term of Eq.~(\ref{eq:psd}) the index is not constrained and the constant absorbs the flat power. In both cases the synthetic light curves are effectively white noise, with little power at multi-year periods. An observed long-period peak is then compared against a background that is too quiet. The spurious bins also increase the variance that normalizes the FTP power, which suppresses the observed peak. The analysis still runs to completion and returns plausible values, and the problem becomes visible only through the dynamic range of the periodogram or through the number of sources crossing the threshold.

\begin{figure*}[t]
    \centering
    \includegraphics[width=0.46\textwidth]{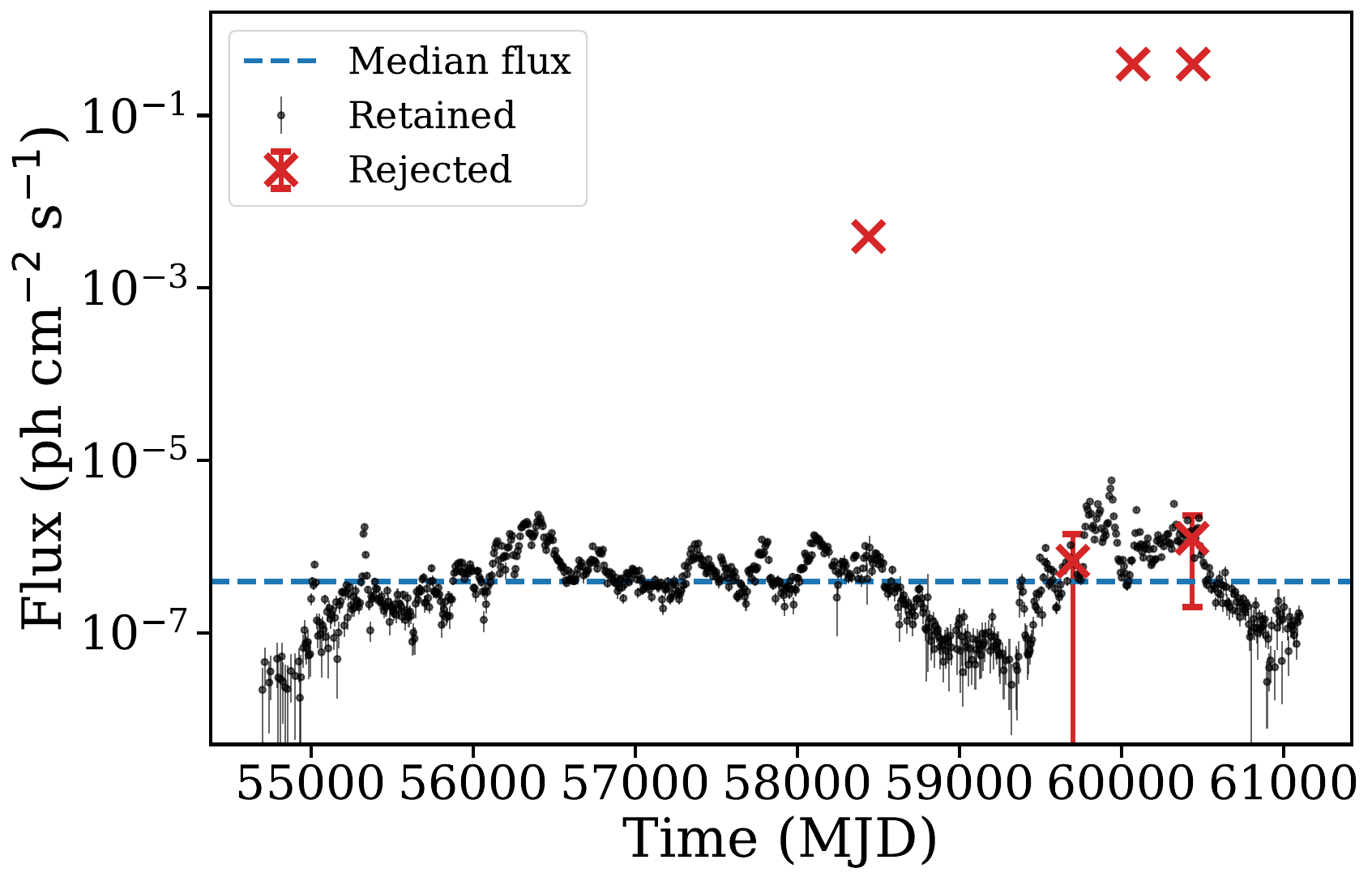}
    \includegraphics[width=0.46\textwidth]{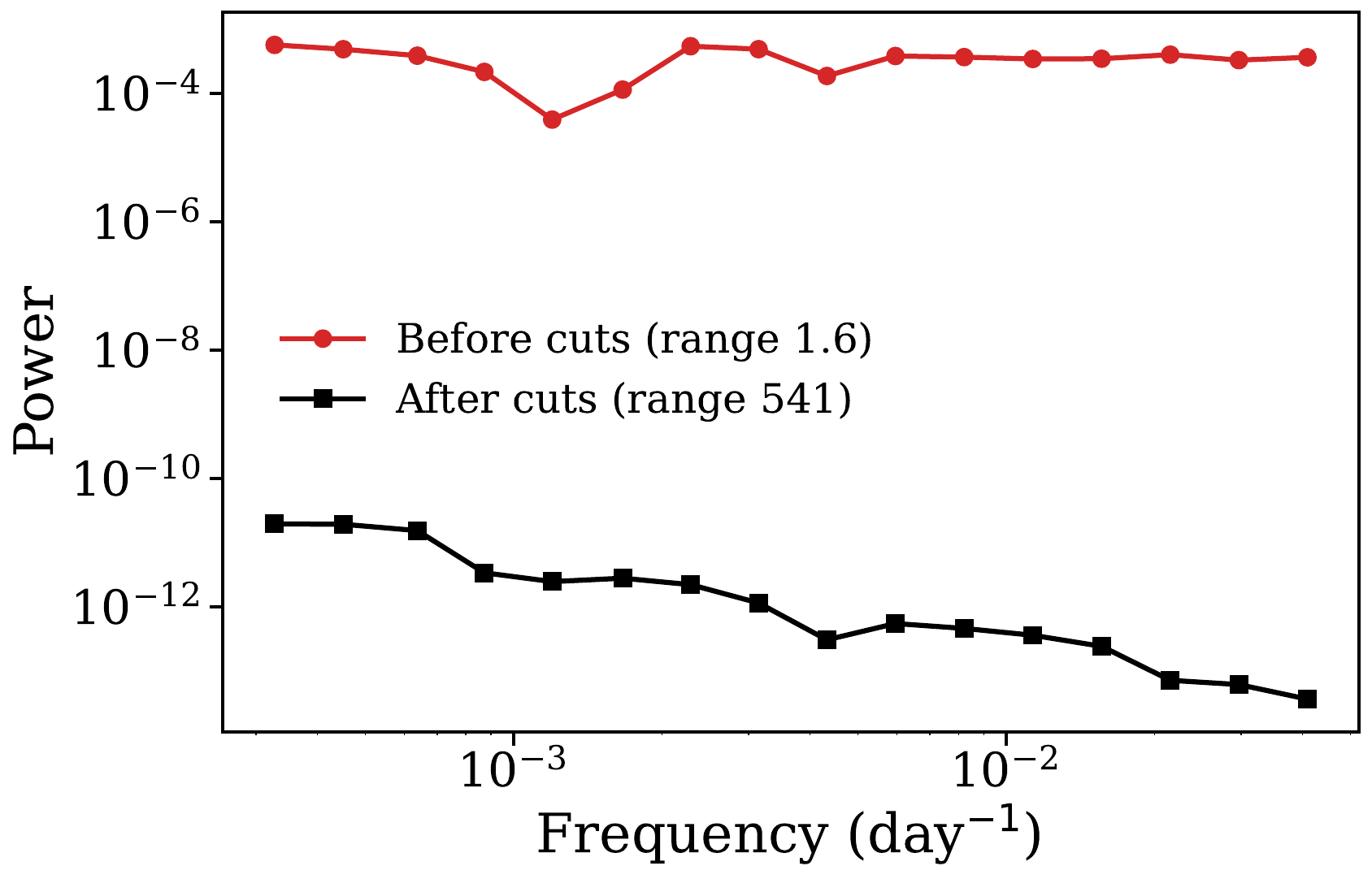}
    \caption{Effect of the quality cuts on 4FGL J1427.9$-$4206 (PKS 1424$-$418), the most affected source in the sample. Left: light curve on a logarithmic flux scale. Bins rejected by the cuts of Sect.~\ref{subsec:sample} are marked with crosses and shown with their uncertainties. Three of them have fluxes four to six orders of magnitude above the median, and the two lying among the retained points have uncertainties comparable to or larger than their flux. Right: binned periodogram before and after the cuts. The dynamic range increases from $1.6$ to $541$.}
    \label{fig:qc_example}
\end{figure*}

Figure~\ref{fig:qc_example} shows the most affected source in the sample, 4FGL J1427.9$-$4206 (PKS 1424$-$418). Before the cuts its periodogram is flat, with a dynamic range of $1.6$, and the index is not constrained ($-0.05 \pm 0.12$ for a pure power law, $3.0 \pm 5.8$ with the constant term). After the cuts, the dynamic range is $541$ and $\alpha = 1.28 \pm 0.08$, a typical value for the $\gamma$-ray light curve of a blazar. The FTP peak powers increase by factors of 10 to 70 across the three templates. With the cuts applied, the highest per-source significance of this source is $1.9\sigma$.

\subsection{Threshold crossings and global significance}
\label{subsec:rate}

In the first pass, with $2\times10^{3}$ realizations, three sources lie within $0.3\sigma$ of the $2.97\sigma$ ceiling, the limit adopted for the follow-up (Sect.~\ref{subsec:fap}), and all three are at the ceiling (Table~\ref{tab:results}, column $\sigma_{\rm 2k}$). These are per-source values, which include the trial factors over periods and over the three templates and are therefore more stringent than a pre-trial local significance of the same nominal value (Sect.~\ref{subsec:fap}). The highest first-pass value among the other 98 sources is $2.61\sigma$ (S5 1044+71). We re-evaluated the three sources with $10^{5}$ realizations, which resolves the values above the ceiling. All three exceed a per-source significance of $3\sigma$: PG 1553+113 ($3.85\sigma$), 4FGL J0102.8+5824 ($3.85\sigma$), and 4FGL J0811.4+0146 ($3.38\sigma$), against $0.14$ expected for the sample. PG 1553+113 is the best-established $\gamma$-ray QPO candidate and is not plausibly a null draw. The other two crossings among the remaining 100 sources correspond to an excess with a Poisson significance of about $2.4\sigma$. Our sample is selected by the completeness of the light curves and is not a random draw. Two sources are too few to tell whether the residual reflects chance, variability not captured by the power-law null, or a remaining calibration issue. None of our conclusions depends on it.

The excess is not confined to the $3\sigma$ tail. Over the whole sample, 14 sources have a first-pass per-source significance above $1.5\sigma$ against 6.7 expected, and 4 above $2.5\sigma$ against 0.6 (Poisson probabilities of 0.010 and 0.004), whereas the numbers above $0.5\sigma$ and $1\sigma$ agree with the expectation. Section~\ref{subsec:dsigma} shows that three independent diagnostics point to the same cause.

No source reaches $3\sigma$ after the correction for the number of sources. The highest values are $2.51\sigma$ for PG 1553+113 and for 4FGL J0102.8+5824, and we report no new periodicity detection. Table~\ref{tab:results} also lists the local significance. For comparison with studies that quote local values, Appendix~\ref{app:local} lists all sources with a local significance of at least $3\sigma$ in the first pass. There are 14, against 12.1 expected from the same rule applied to the synthetic light curves, with a probability of 0.32 of finding at least as many. The number of local $3\sigma$ crossings is therefore what the null predicts, and the sources in that table are not individually evidence of periodicity.

\input{table1}

\subsection{Dependence of the significance on the template}
\label{subsec:dsigma}

We quantify the effect of the template with $\Delta\sigma = \sigma_{\text{best}} - \sigma_{\text{sine}}$, the difference between the per-source significance of the best-fitting template and that of the sinusoid. Because both values are obtained for the same source, detrending, and realizations, and differ only in the template, $\Delta\sigma$ isolates the dependence on the assumed morphology within the adopted calibration. Because $\sigma_{\text{best}}$ is the maximum over the three templates, $\Delta\sigma \geq 0$ by construction, and the null distribution below includes this selection.

\begin{figure*}[t]
    \centering
    \includegraphics[width=0.33\textwidth]{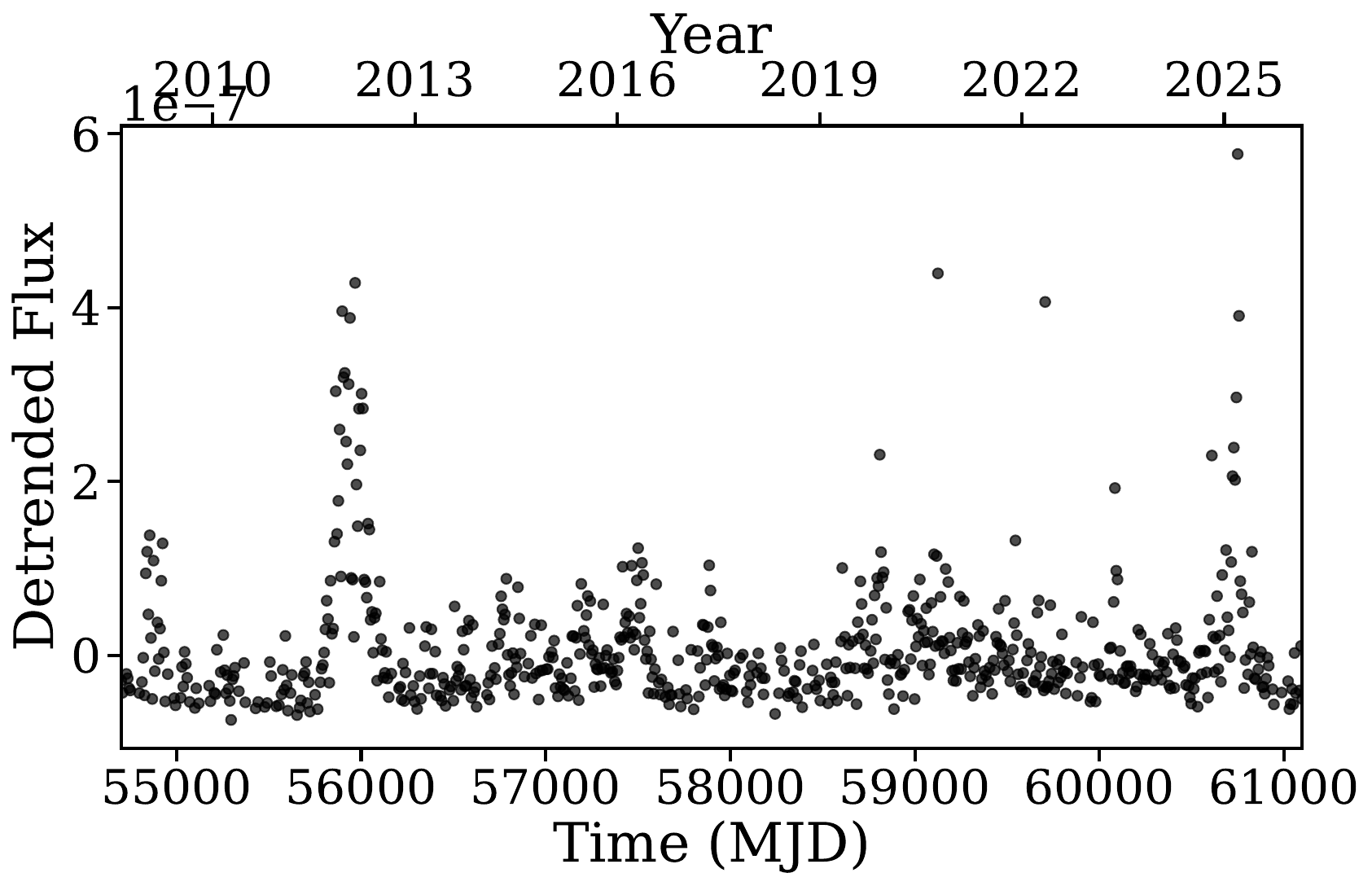}
    \includegraphics[width=0.33\textwidth]{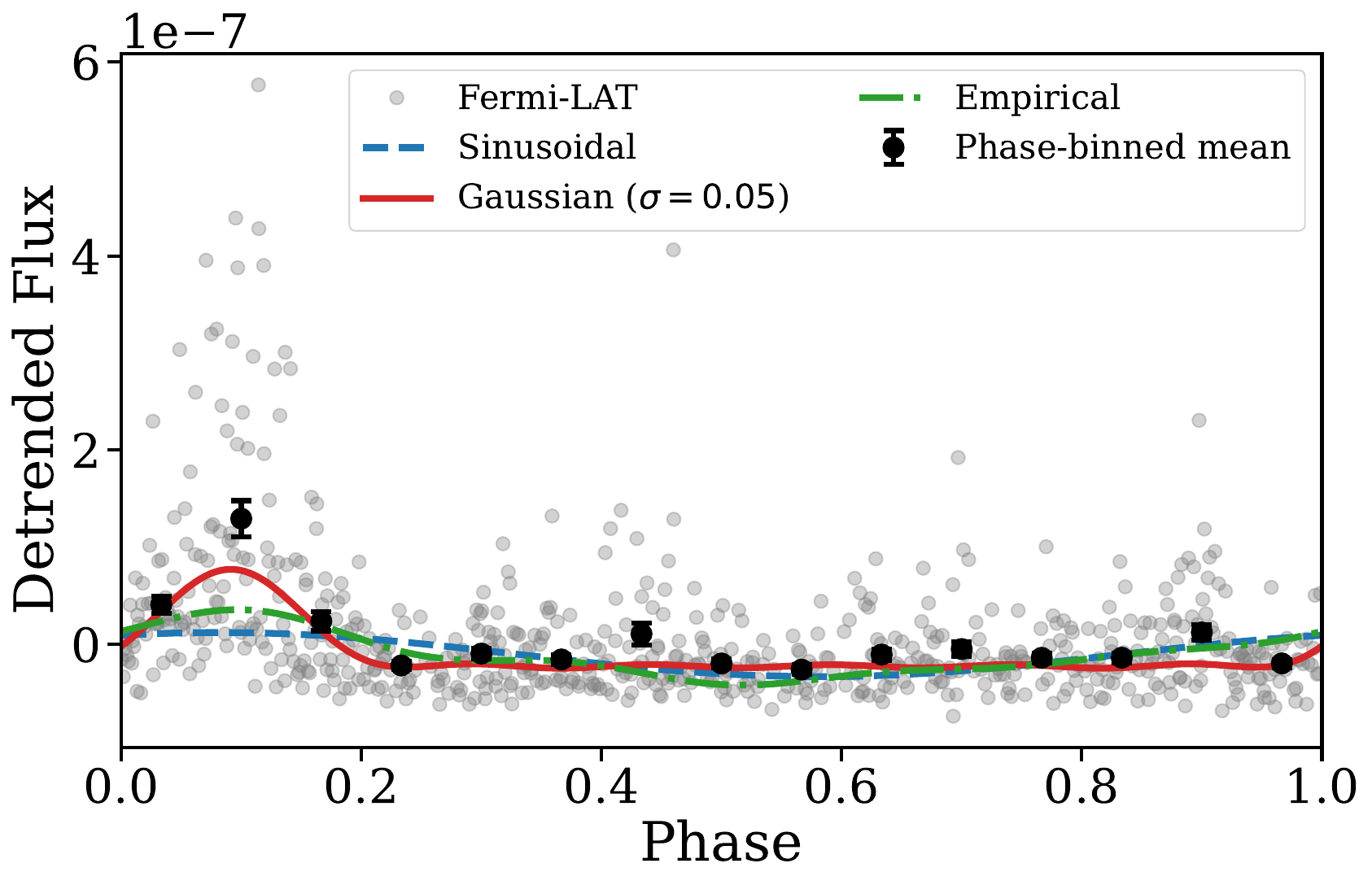}
    \includegraphics[width=0.33\textwidth]{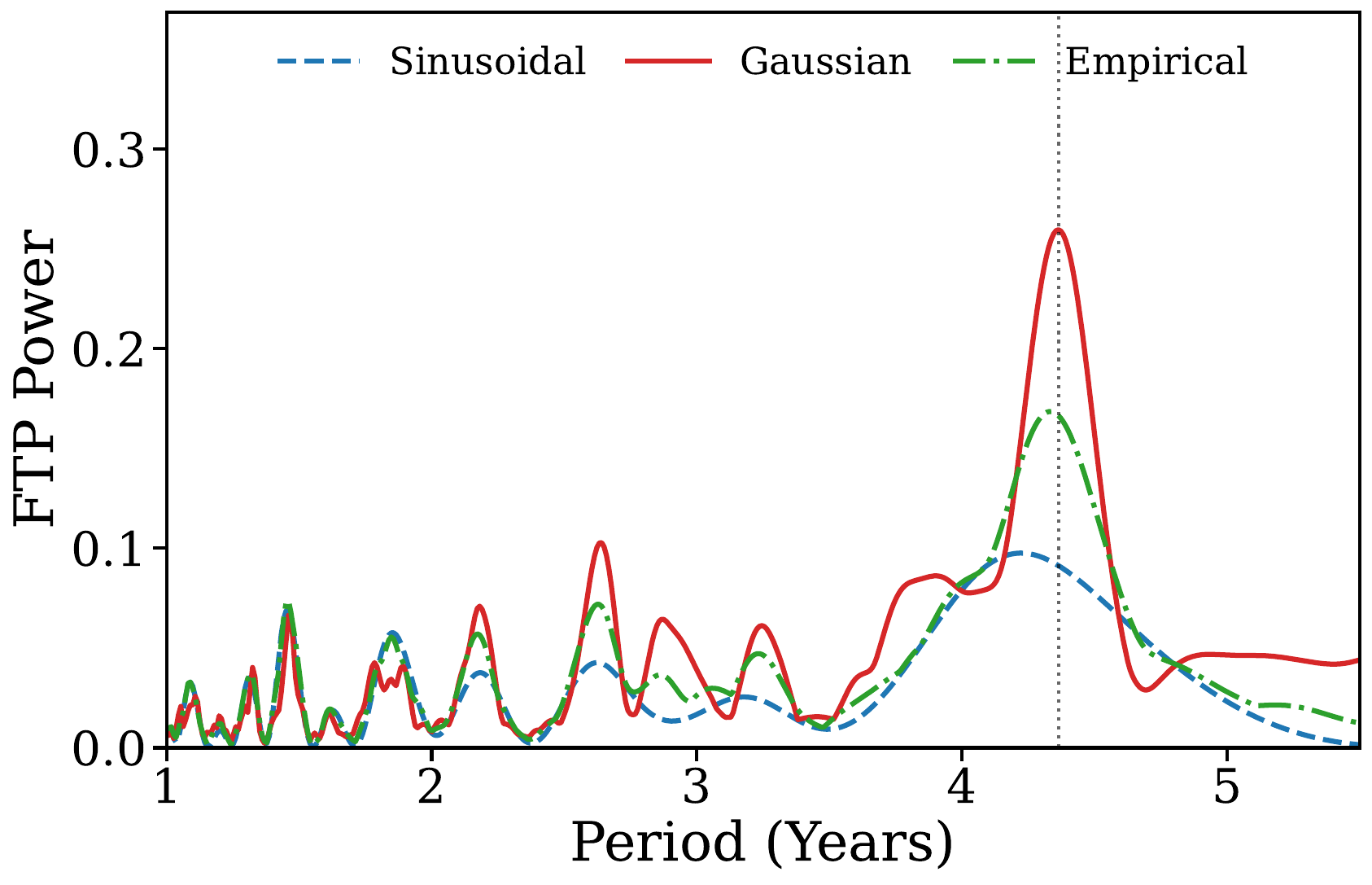}\\[0.5ex]
    \includegraphics[width=0.33\textwidth]{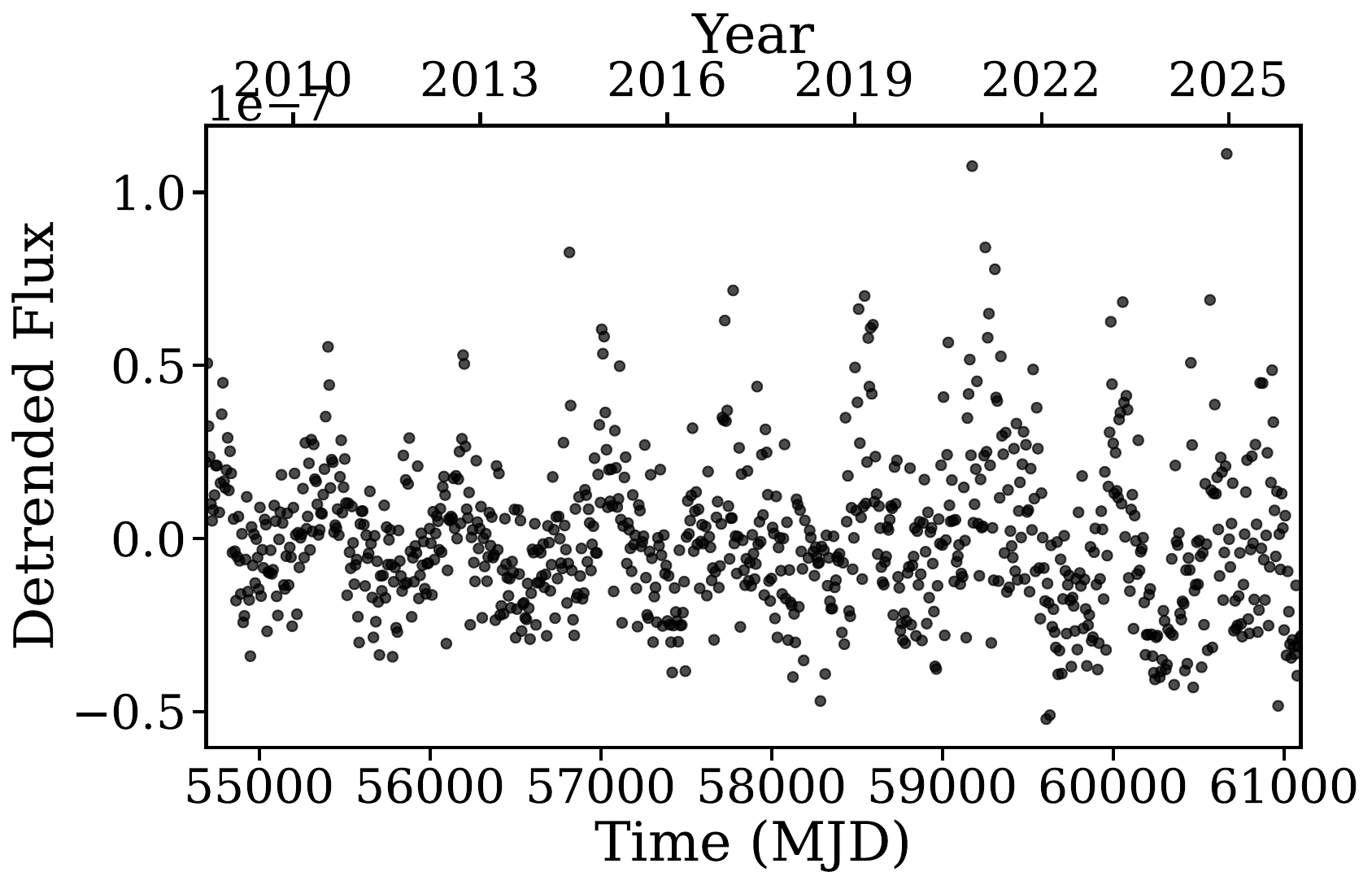}
    \includegraphics[width=0.33\textwidth]{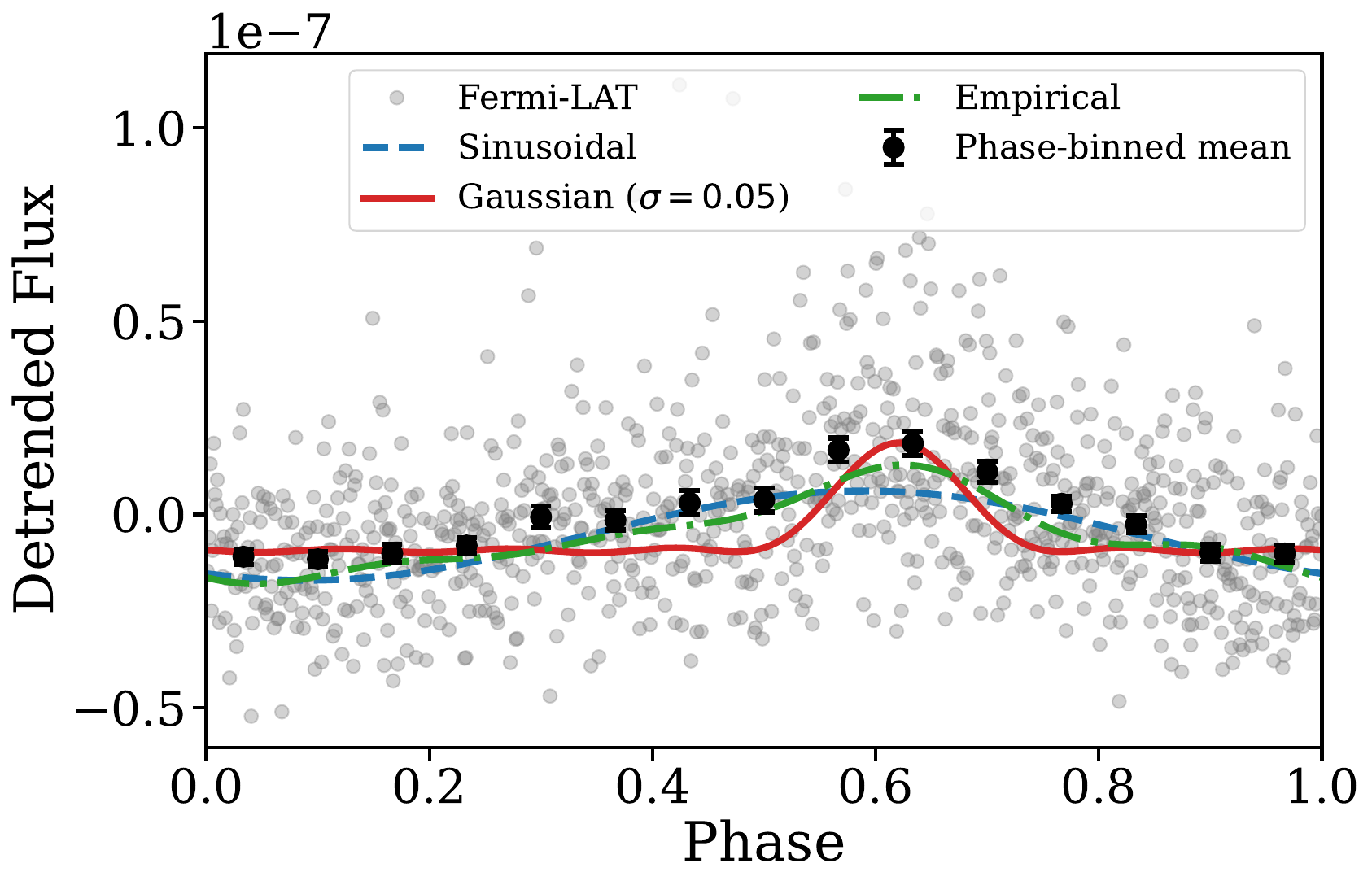}
    \includegraphics[width=0.33\textwidth]{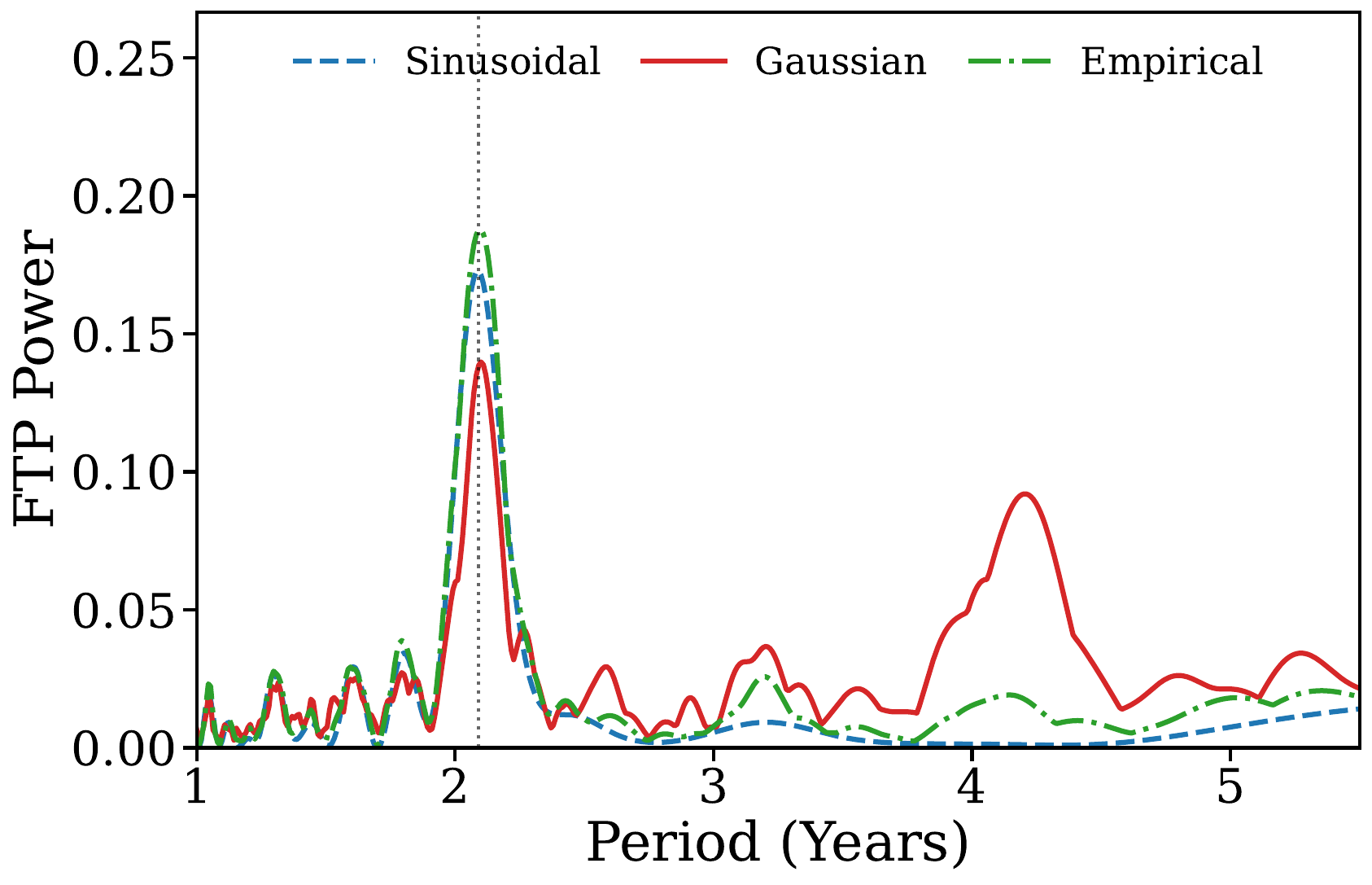}\\[0.5ex]
    \includegraphics[width=0.33\textwidth]{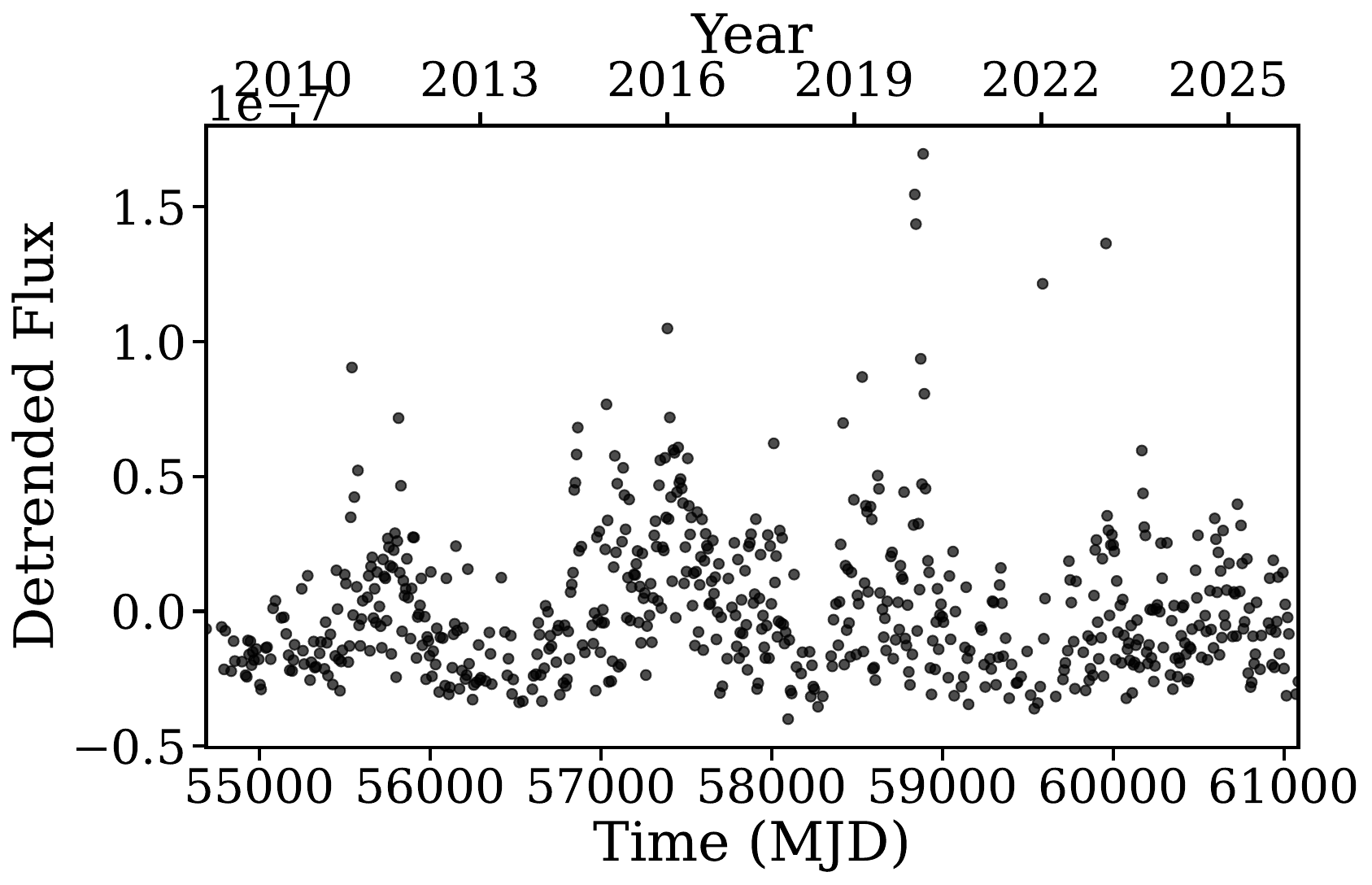}
    \includegraphics[width=0.33\textwidth]{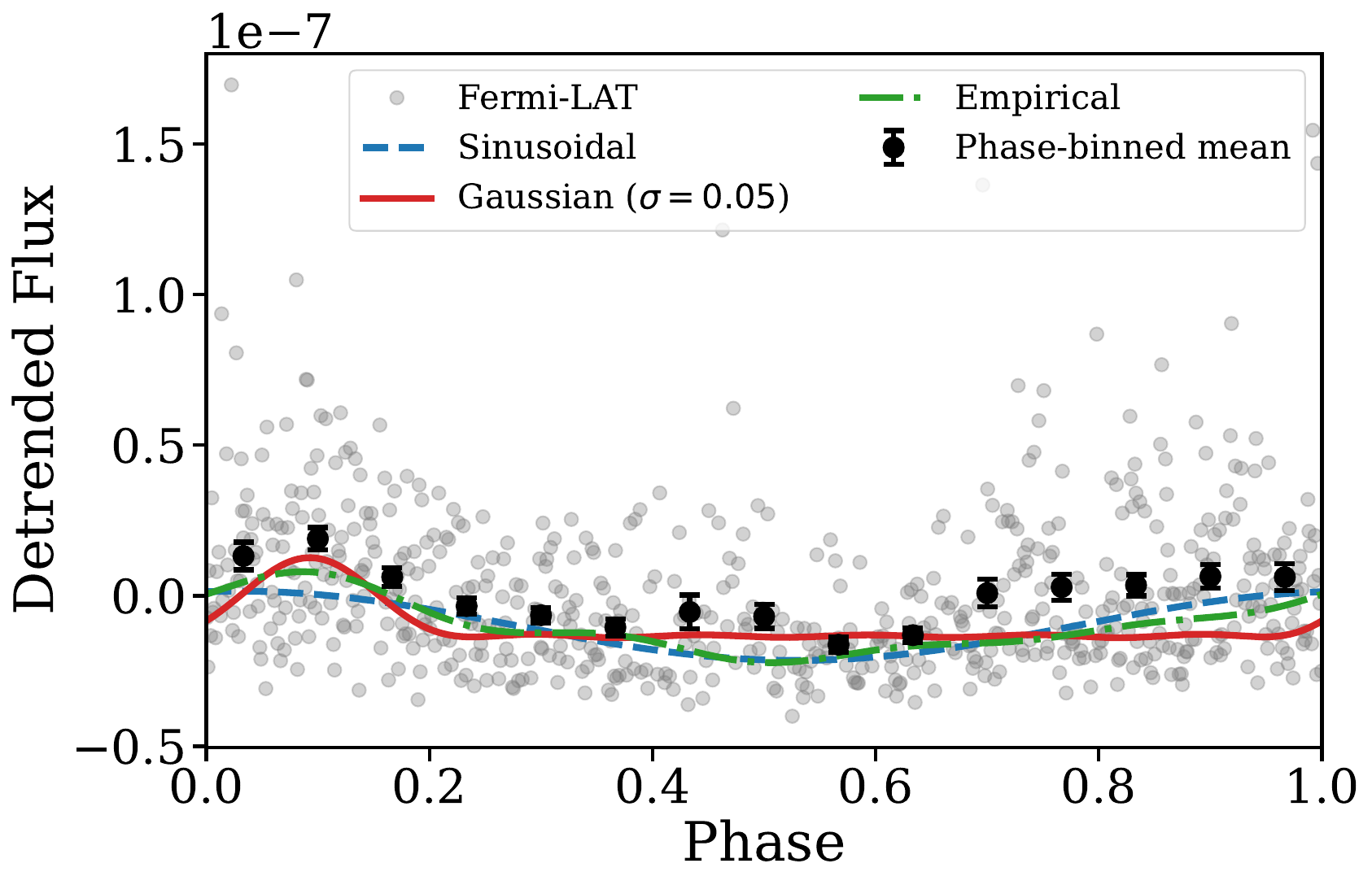}
    \includegraphics[width=0.33\textwidth]{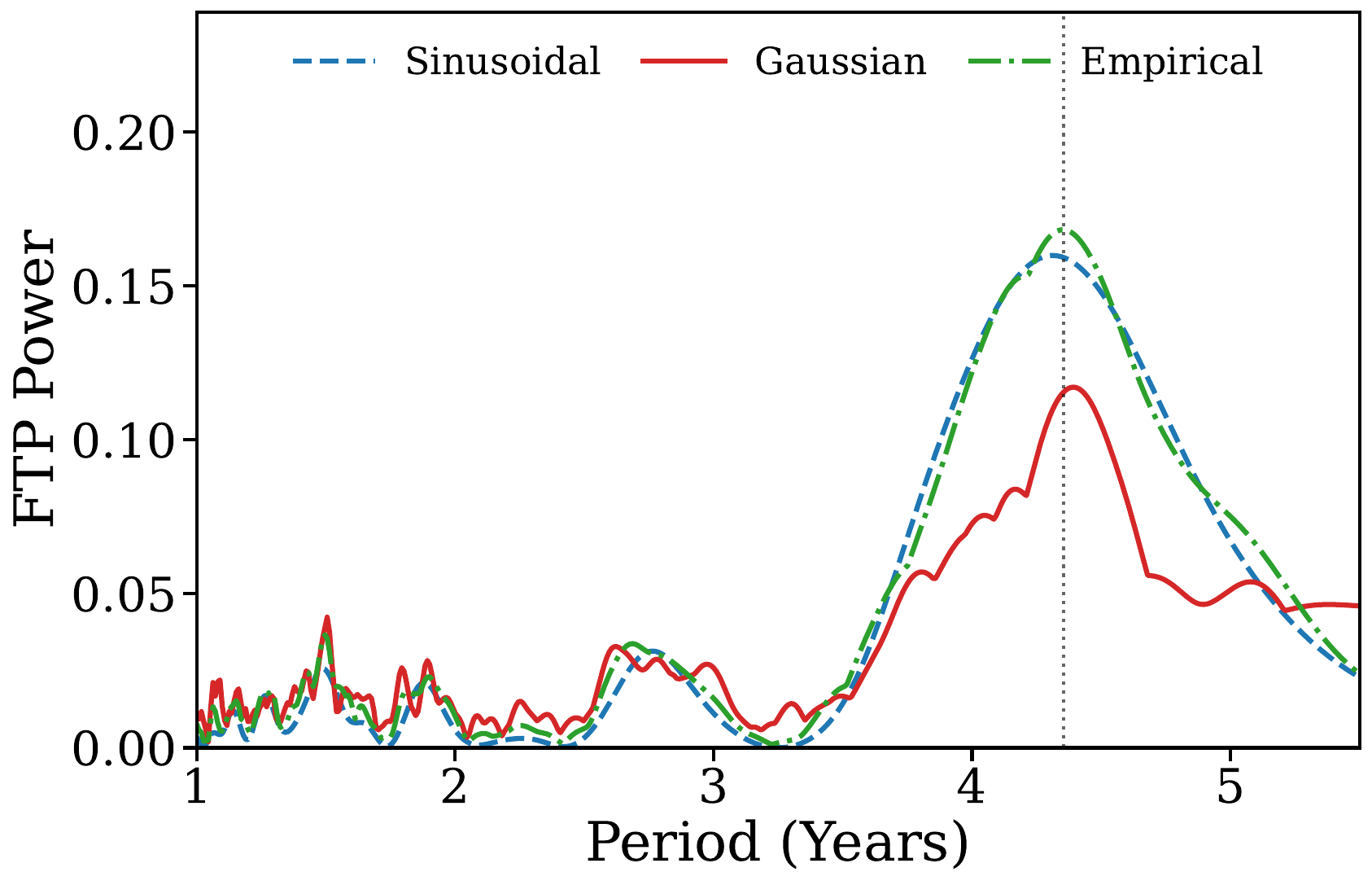}
    \caption{Analysis of the three sources in Table~\ref{tab:results}: 4FGL J0102.8+5824 (TXS 0059+581, top), 4FGL J1555.7+1111 (PG 1553+113, middle), and 4FGL J0811.4+0146 (OJ 014, bottom). Left: SSA-detrended light curve. Center: phase-folded light curve with the phase-binned mean (black) and the best-fitting model of each template, folded at the period of the template with the highest per-source significance. Right: FTPs for the three templates. For 4FGL J0102.8+5824 the per-source significance is $0.24\sigma$ with the sinusoid and $3.85\sigma$ with the $\sigma = 0.05$ Gaussian burst.}
    \label{fig:j0102_analysis}
\end{figure*}

The most pronounced case among the sources evaluated with $10^{5}$ realizations is 4FGL J0102.8+5824 (TXS 0059+581), shown in Fig.~\ref{fig:j0102_analysis}. Its per-source significance is $0.24\sigma$ with the sinusoidal template and $3.85\sigma$ with the Gaussian burst ($\Delta\sigma = 3.61$). This is consistent with a small number of multi-month active states, which a single narrow burst per cycle matches better than a sinusoid. Its global significance is $2.51\sigma$.

\begin{figure*}[t]
    \centering
    \includegraphics[width=0.46\textwidth]{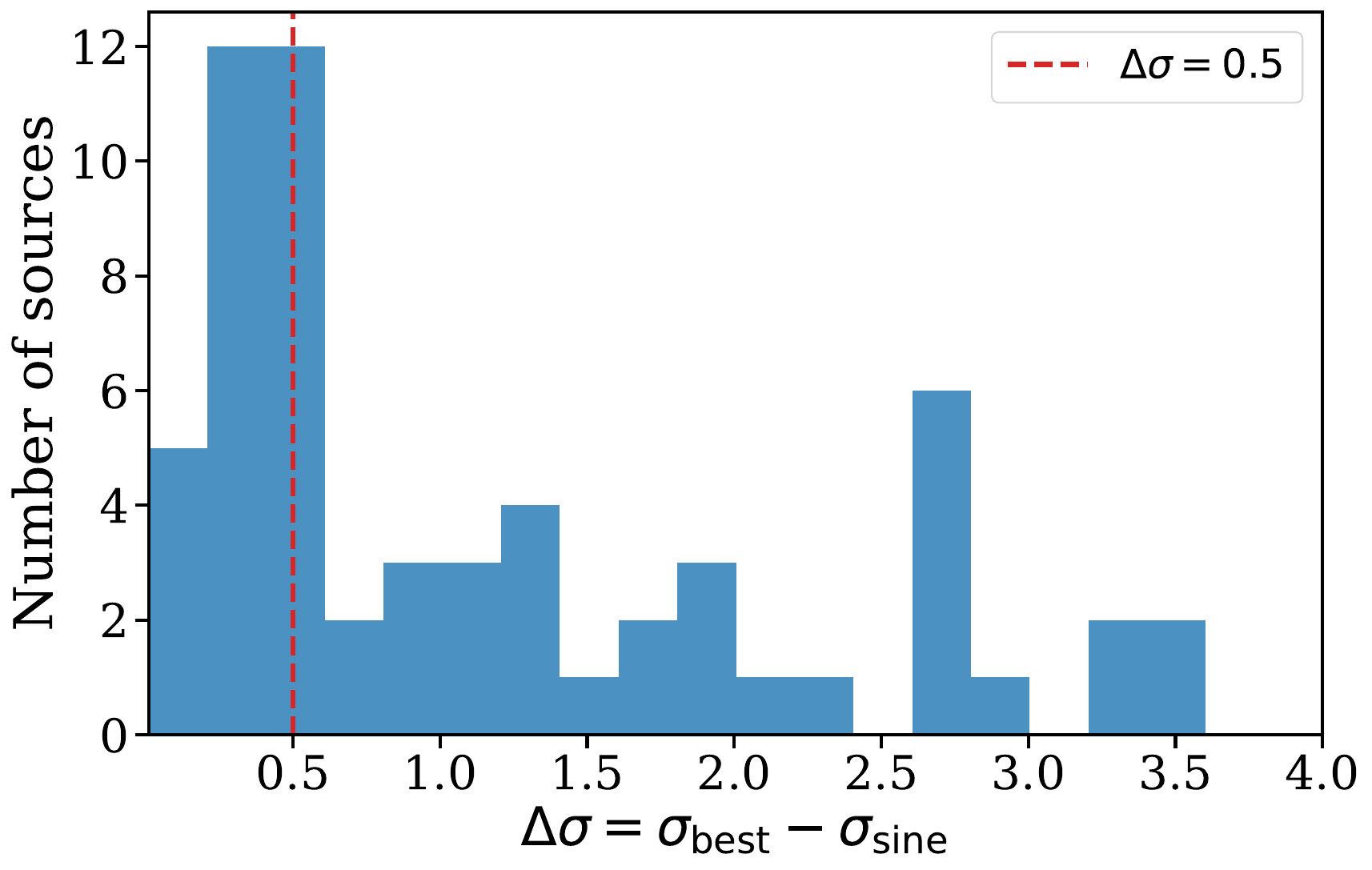}
    \includegraphics[width=0.46\textwidth]{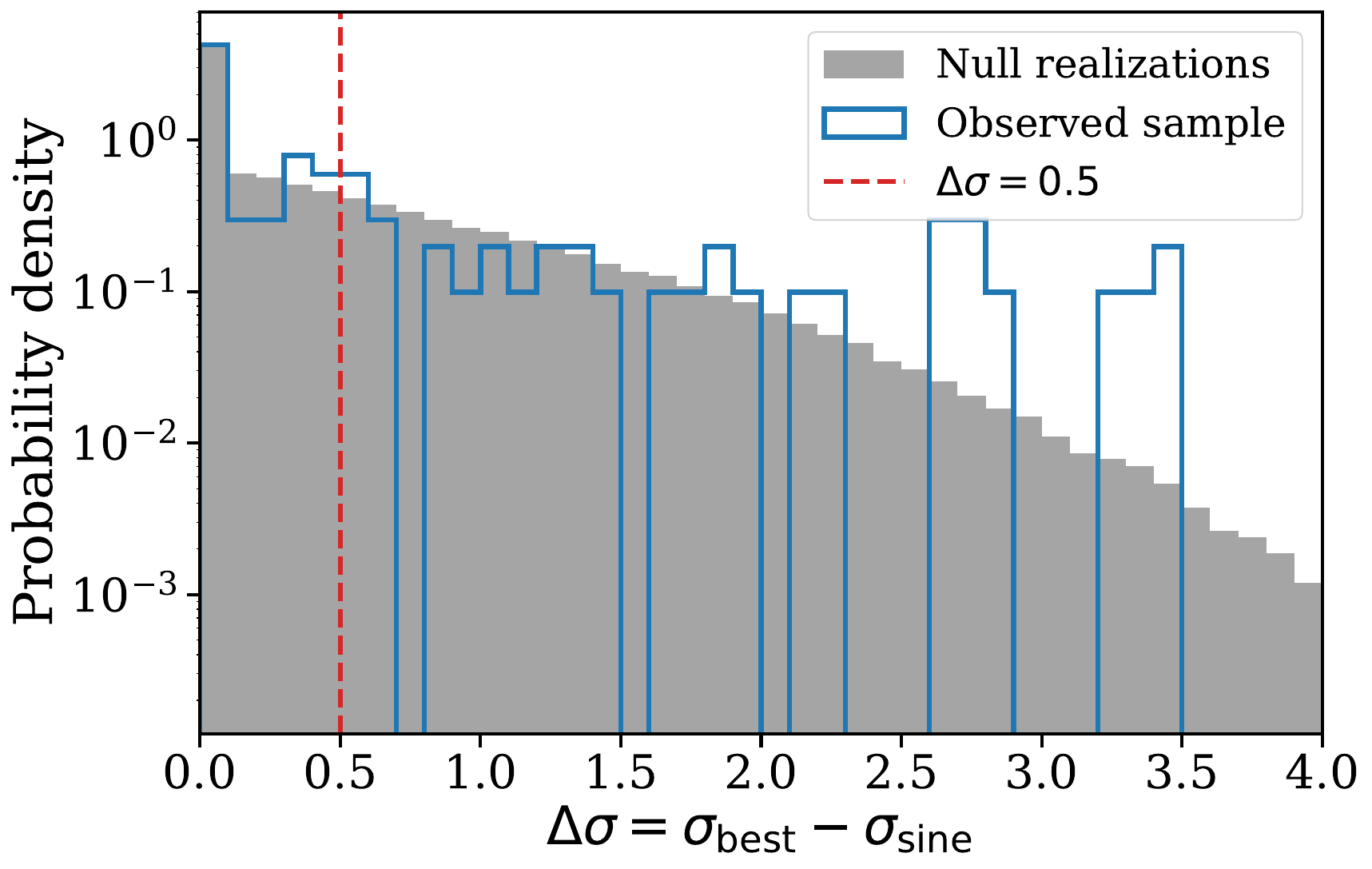}
    \caption{Distribution of $\Delta\sigma = \sigma_{\rm best} - \sigma_{\rm sine}$. Left: the 101 sources of the sample. The axis begins at $0.01$ and excludes the 41 sources with $\Delta\sigma = 0$. Right: the observed sample (blue) compared with $2.0\times10^{5}$ null realizations (gray), the first $2\times10^{3}$ of each source. The axis ends at $\Delta\sigma = 4$, beyond which there are no observed values and 0.06\% of the null realizations. Both are evaluated at the resolution of the first pass. The dashed line marks $\Delta\sigma = 0.5$.}
    \label{fig:dsigma}
\end{figure*}

We evaluate $\Delta\sigma$ for the whole sample at the resolution of the first pass ($2\times10^{3}$ realizations per source). In these terms, 41 of the 101 sources have $\Delta\sigma = 0$, meaning that the sinusoid is the best-fitting template, and 38 have $\Delta\sigma > 0.5$ (Fig.~\ref{fig:dsigma}, left). To determine how $\Delta\sigma$ behaves without a signal, we computed it at the same resolution for each of the first $2\times10^{3}$ synthetic light curves of every source, $2.0\times10^{5}$ in total, taking the FAP of each realization against the remaining realizations of the same source. Under the null, $\Delta\sigma$ exceeds $0.5$ in 37\% of the realizations, $1$ in 20\%, $2$ in 4.3\%, and $3$ in 0.6\%. A value of $\Delta\sigma = 0.5$ is therefore not evidence of a non-sinusoidal morphology. At the first-pass resolution, 4FGL J0102.8+5824 has $\Delta\sigma = 2.69$, which is reached by 1.1\% of the null realizations, and $1.1$ sources are expected to exceed it in a sample of 101. The values obtained with $10^{5}$ realizations cannot be compared with this null, because first-pass significances are limited to $2.97\sigma$, and we do not quote a percentile for them.

The observed distribution agrees with the null below $\Delta\sigma \approx 1$ but has a heavier tail: 13 sources exceed $\Delta\sigma = 2$, against 4.3 expected (Poisson probability $6\times10^{-4}$), and 4 exceed $\Delta\sigma = 3$, against 0.6 expected ($3\times10^{-3}$; Fig.~\ref{fig:dsigma}, right). In 12 of these 13 sources the sinusoidal significance is negative, and the large $\Delta\sigma$ reflects a weak response to the sinusoid more than a strong response to another template. Among the sources in Table~\ref{tab:results}, only 4FGL J0102.8+5824 combines a large $\Delta\sigma$ with a best-template significance above $3\sigma$. A heavier tail is expected if the light curves contain isolated flares absent from the Gaussian red-noise null, because a narrow burst template responds to such flares whether or not they recur. Flares are known to distort periodicity searches, changing the recovered period and significance by more than 100\% in some cases \citep{penil2025d}. That study injected a flare into a periodic signal, whereas here the flares are absent from the null. The tail is consistent with light curves that are more burst-dominated than the null, but it does not indicate periodicity.

Two further diagnostics point the same way. First, the single-template FAPs have an excess of small values for all three templates: the number of sources with a FAP below 1\% is 5 for the sinusoid, 4 for the empirical template, and 7 for the Gaussian burst, against 1.0 expected (Poisson probabilities of 0.004, 0.02, and $9\times10^{-5}$). The Gaussian burst, which responds most strongly to isolated flares, is the most affected, and the significances it returns for burst-like light curves, including the $3.85\sigma$ of 4FGL J0102.8+5824, may therefore be overestimated. Second, the classifier score responds to simulated isolated flares (Sect.~\ref{subsec:ml_results}).

\subsection{Structural ranking}
\label{subsec:ml_results}

\begin{figure*}[t]
    \centering
    \includegraphics[width=0.49\textwidth]{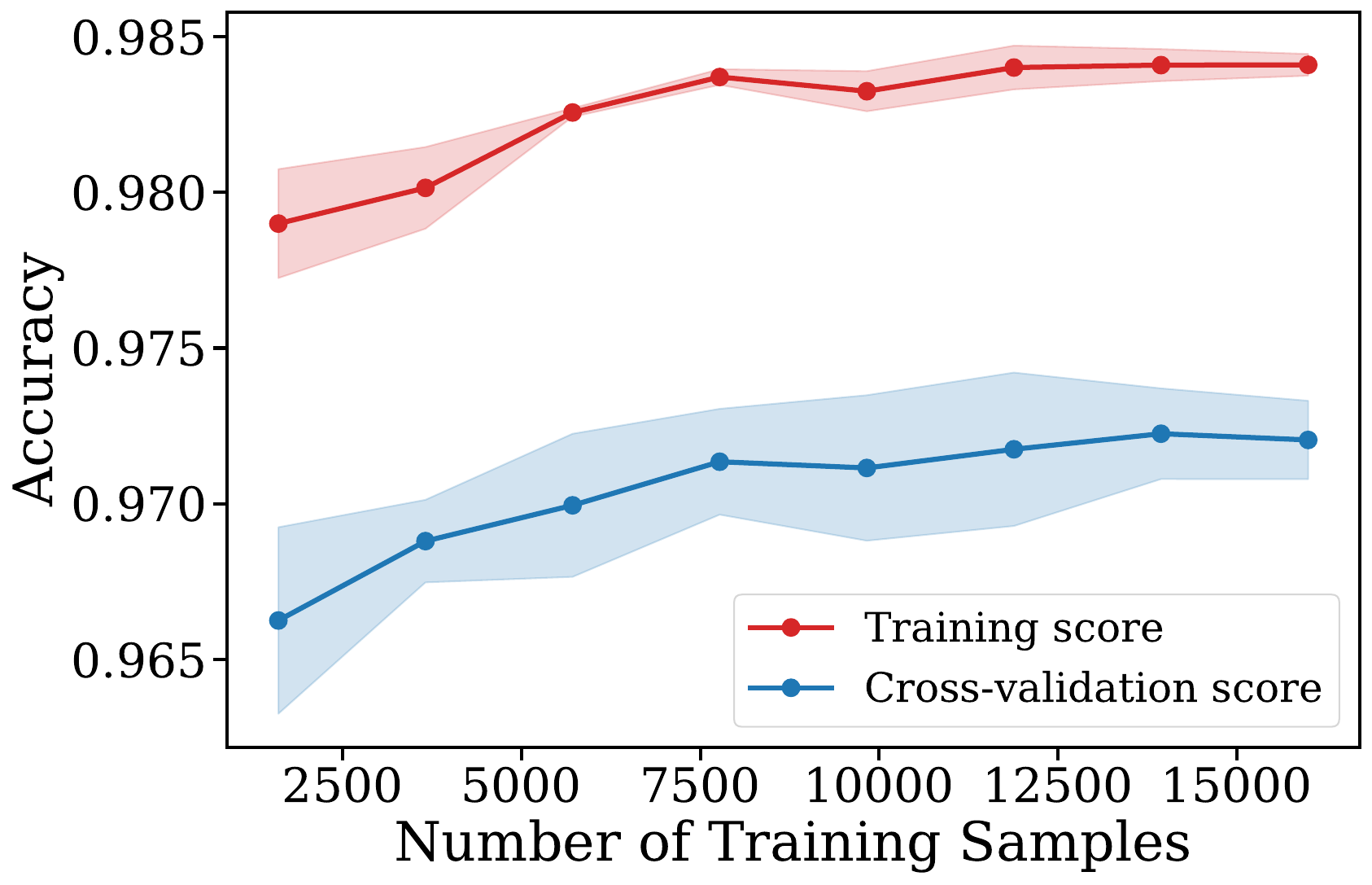}
    \includegraphics[width=0.49\textwidth]{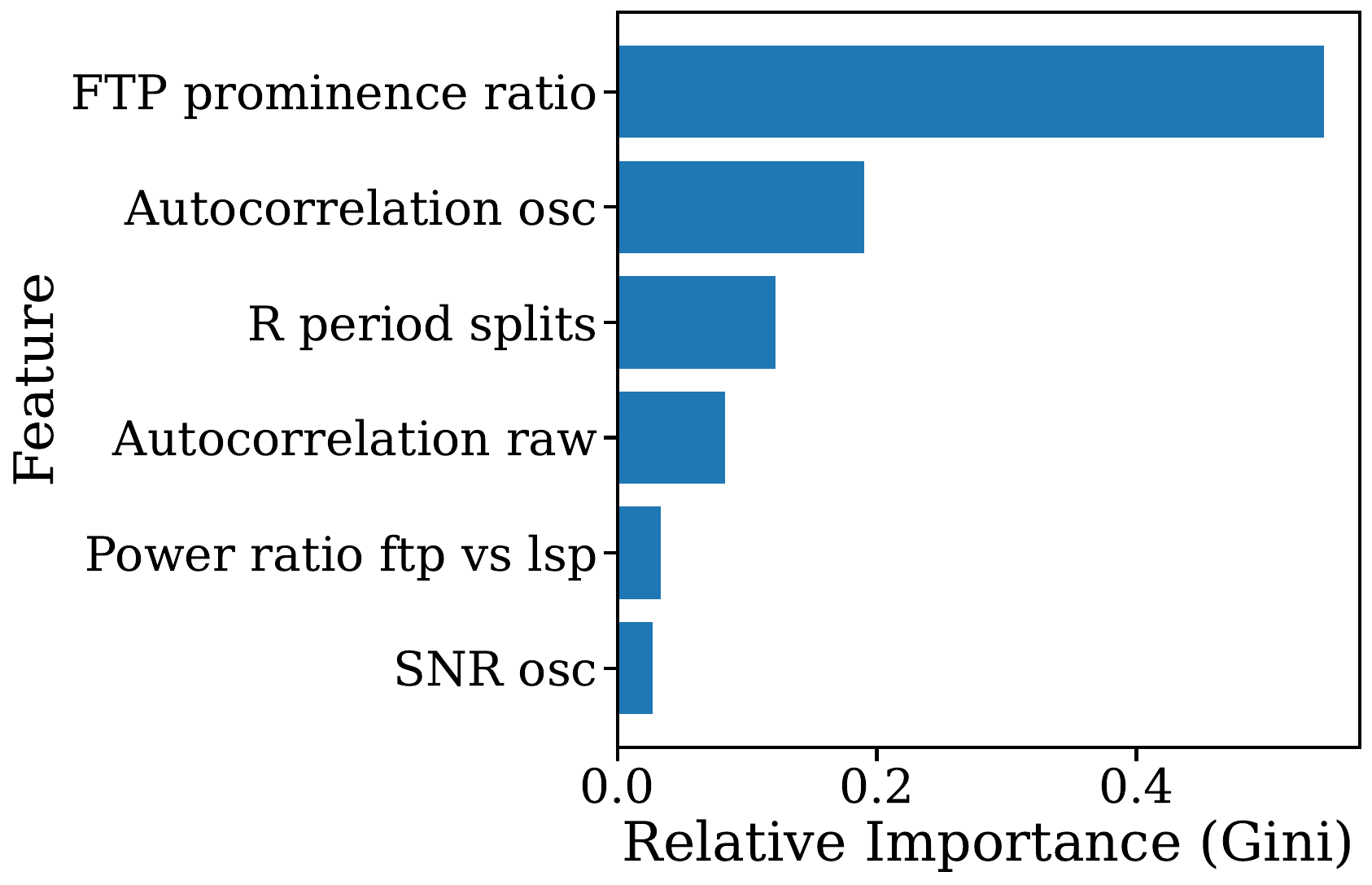}
    \caption{Random Forest classifier. Left: learning curve on the synthetic training set. The plateau shows that performance is not limited by the size of the training set. It is an internal diagnostic and does not establish how the classifier performs on observed light curves, for which no labeled validation set exists. Right: Gini feature importances.}
    \label{fig:ml_performance}
\end{figure*}

The classifier reaches a cross-validated accuracy of $0.972 \pm 0.001$ on the synthetic training set (Fig.~\ref{fig:ml_performance}). The prominence of the periodogram peak carries 54\% of the Gini importance, and the autocorrelation of the oscillatory component (19\%), the split-half period ratio (12\%), and the autocorrelation of the raw light curve (8\%) carry most of the remainder.

We evaluated $S_{\rm morph}$ for all 101 sources. The median score is $0.03$, and ten sources score above $0.5$. The scores of the sources in Table~\ref{tab:results} are $0.81$ for 4FGL J0102.8+5824, $0.79$ for PG 1553+113, and $0.11$ for 4FGL J0811.4+0146. The highest scores belong to S5 1044+71 ($0.99$), S4 1144+40 ($0.98$), and 3C 371 ($0.96$). The score is only moderately related to the per-source significance (Spearman coefficient $0.46$), and it is high for sources of low significance: 3C 371 has a per-source significance of $2.3\sigma$ and a global significance below $1\sigma$, and PKS 1424$-$418 has $0.86$ for $1.9\sigma$. The light curves, phase-folded profiles, and periodograms of the three sources in Table~\ref{tab:results} are shown in Fig.~\ref{fig:j0102_analysis}.

To test what the score responds to, we applied the trained classifier to 2000 synthetic light curves of each of four kinds, generated like the training set with the gaps, uncertainties, and PSD parameters of the sources. For Gaussian red noise, 1.7\% of the light curves score above $0.5$. For skewed noise, obtained as the exponential of the Gaussian realization, the fraction is 1.2\%. For Gaussian red noise with non-recurring flares, six on average per light curve, with exponential rise and decay times of 7--28 and 28--140~d and amplitudes drawn from an exponential distribution with a mean of three times the noise standard deviation, the fraction is 8.4\%, and 4.2\% score above $0.65$. Among the observed sources, 10 score above $0.5$ and 9 above $0.65$. These numbers are far above the expectation for Gaussian noise (1.7 sources, probability $7\times10^{-6}$) and consistent with that for noise with flares (8.4 sources, probability $0.34$). As a positive control, periodic signals with $50\%$ cycle-to-cycle amplitude scatter and $0.05$ cycle of phase jitter on Gaussian noise score above $0.5$ in 92\% of the cases. The score therefore recognizes periodic signals in simulations, but it also responds to isolated flares, and in the observed sample it cannot distinguish the two. The flare model is a simple choice, and the agreement with the observed count is indicative.

\section{Discussion}
\label{sec:discussion}

\subsection{Data quality in periodicity searches}
\label{subsec:dq}

The LCR documentation \citep{abdollahi2023} advises against using non-convergent fits in higher-level analyses, and Sect.~\ref{subsec:qc} shows how large the effect of retaining them can be in a periodicity search. The consistency test on $F/\sqrt{\text{TS}}$ complements that criterion for the few bins whose fits converged but are poorly constrained. Two inexpensive diagnostics defined in Sect.~\ref{subsec:fap} reveal the problem: the dynamic range of the binned periodogram, and the comparison between the number of sources above a threshold and the number expected by chance. Both are simple to report in periodicity searches based on LCR light curves. When independent analyses recover compatible periods for a source but disagree on its significance, the preprocessing of the light curves is a plausible cause, besides the timing method.

The cuts matter more for 7-day than for 30-day binning. For the same 101 sources, 0.41\% of the 7-day bins have non-convergent fits, against 0.04\% of the 30-day bins. The fraction is small, but the rejected bins are extreme and concentrated in a few light curves, and their effect on a periodicity search is not proportional to their number: 10 of the 907 bins of PKS 1424$-$418 are enough to flatten its periodogram (Sect.~\ref{subsec:qc}), and without the cuts 22 of the 101 light curves have a flat periodogram, 19 of which become red once the cuts are applied. The three cuts reject 135 of 77\,393 bins with $\text{TS} \geq 2$ at 7 days (0.17\%), affecting 52 sources, and 12 of 21\,005 bins at 30 days (0.06\%), affecting 11 sources. Shorter bins contain fewer photons, which makes unconstrained fits more frequent, and \citet{penil2026b} find that the relative flux uncertainty of 7-day bins is about twice that of monthly bins. Searches that use monthly binning, as the studies discussed in Sect.~\ref{subsec:lit} typically do, are therefore less exposed to the problem described here. Weekly binning has so far been used in systematic searches of samples of bright sources \citep{ren2023, chen2024}.

\subsection{Comparison with previous work}
\label{subsec:lit}

Our modest yield and low global significances agree with recent systematic searches. \citet{penil2025c} report a global significance near $1.8\sigma$ for PG 1553+113, and several previously reported candidates fall to values consistent with zero once the look-elsewhere effect is taken into account. A search over 1492 jetted AGN finds no statistically significant periodic population \citep{penil2025a}, and \citet{rico2025} reach comparable conclusions for several hundred sources. \citet{penil2026} revisit periodicity candidates with more than 17 years of data and several stochastic models, and find global significances consistent with zero for sources with local significances above $3\sigma$, among them PG 1553+113 and S5 1044+71. Our per-source significances include the trial factors over periods and templates and cannot be compared directly with the pre-trial local significances quoted in these studies. Global values depend on the number of sources searched, 101 here against 494 and 1492 in those works, and we therefore compare periods and not significances. Those searches typically use monthly binning and evaluate the periodogram of the SSA oscillatory component \citep[e.g.,][]{rico2025}, which suppresses short-timescale variability. Our 7-day binning preserves intra-cycle structure but retains that variability, against which any periodic signal must be detected.

Table~\ref{tab:local} lists, for the 14 sources with a local significance of at least $3\sigma$, the periods and significances published by \citet{rico2025} and \citet{penil2026}; nine of them appear in at least one of those works. We compare periods only, because the significances are not equivalent quantities. The local significance of \citet{rico2025} includes the search over periods and comes from 12 yr of 28-day light curves. The value quoted by \citet{penil2026} is a median over methods and noise models, which those authors use only to sort the sources, obtained from 17.3 yr of 30-day light curves. Ours is obtained with 7-day binning, three templates, and a power-law null. With a tolerance of 15\% on the published period, our period agrees with at least one published period for all nine sources (7 of 9 for \citealt{rico2025} and 8 of 8 for \citealt{penil2026}). The test is lenient for sources with two published periods, and since all three analyses use largely the same photons, the agreement is not independent evidence of periodicity.

All three sources in Table~\ref{tab:results} have been examined before. For PG 1553+113 we recover a period of $2.09$~yr, consistent with the $2.18 \pm 0.08$~yr of \citet{ackermann2015} and with the reanalysis of \citet{penil2026}, who find the period supported by several methods. The 2.1-year oscillation of PG 1553+113 has also been described as a broad activity envelope with shorter-timescale substructure, not as a strictly sinusoidal modulation \citep{penil2026b}. \citet{penil2026} list 4FGL J0811.4+0146 (OJ 014) among the sources with a local significance of at least $2\sigma$ ($2.5\sigma$, period $4.2$~yr), and we recover $4.3$~yr with a per-source significance of $3.38\sigma$ and a global significance of $1.80\sigma$. Among the sources of Table~\ref{tab:local}, S5 1044+71 is the other source for which \citet{penil2026} find a local significance above $3\sigma$. We recover $3.17$~yr, consistent with the $3.06 \pm 0.43$~yr of \citet{wang2022}. \citet{ren2023} and \citet{penil2026} also recover this period, and \citet{cerruti2026} model it as a precessing jet. Its per-source significance in our analysis is $2.61\sigma$, and its global significance is below $1\sigma$. No $\gamma$-ray periodicity has been reported for 3C 371 to our knowledge, and its per-source significance of $2.3\sigma$ does not suggest one, although its classifier score is $0.96$ (Sect.~\ref{subsec:ml_results}).

A period of 4.0~yr was reported for 4FGL J0102.8+5824 (TXS 0059+581) by \citet{penil2025c}, with an average local significance of $0.8\sigma$ over their methods. In a study of the effect of flares on periodicity searches, \citet{penil2025d} find a different period for this source ($1.8 \pm 0.2$~yr at $1.2\sigma$ with SSA), identify a prominent flare in the middle of its light curve, and conclude that the previously suggested periodic behavior is not genuine. With 17.6 yr of data, the 4.4-year peak of the Gaussian burst template in our analysis does not depend on a single flare. Removing the three brightest flares lowers its power from $0.26$ to $0.19$ without moving it, and in each of the four cycles the brightest bin falls at phase 0.77 to 0.80, including a flare in 2025 that is later than the data analyzed by \citet{penil2025c, penil2025d}. These flares define the period, however, and four cycles are few. The narrow Gaussian burst responds most strongly to a few discrete, well-separated flares and is the template most affected by flares absent from the null (Sect.~\ref{subsec:dsigma}), a gain of this size is expected for about one source in our sample under the null, and the global significance is $2.51\sigma$. We therefore do not regard this source as a periodicity candidate. It illustrates both the sensitivity of the significance to the template and the fact that a large gain from a non-sinusoidal template is not evidence of periodic structure.

\subsection{The role of the structural classifier}
\label{subsec:mlrole}

The FAP is an amplitude statistic: it measures how often a stochastic process produces a peak of a given height in the searched band, not whether a periodicity persists across the baseline. The split-half period ratio and the autocorrelation features measure persistence without reference to the amplitude of any template fit, and together they carry about 39\% of the importance of the classifier. Two sources with the same FAP can therefore receive different scores. The score and the per-source significance are only moderately related (Spearman coefficient $0.46$), which is expected because both respond to the height of the periodogram peak.

The score is not calibrated against a population of realistic signals and noise, and a value near unity cannot be read as a probability. Because the noise class of the training set is Gaussian, the score also responds to isolated flares (Sect.~\ref{subsec:ml_results}): the number of observed sources scoring above $0.5$ is consistent with that expected for red noise with non-recurring flares and far above that for Gaussian noise. High scores, such as those of 3C 371 and PKS 1424$-$418, therefore do not indicate periodicity, and the score has no absolute interpretation. We keep it as a ranking statistic for the comparison with the FAP and do not use it to select sources.

The practical lesson concerns the training set. A classifier trained against Gaussian red noise inherits the blind spot of that null: anything the null lacks, such as isolated flares, can look like signal, and flares can also suppress a genuine periodic signal \citep{penil2025d}. Its noise class therefore needs the same realism as the significance calibration. In our sample, the main contribution of the classifier is a third indication, together with the tail of $\Delta\sigma$ and the excess of small FAPs (Sect.~\ref{subsec:dsigma}), that the light curves contain flares not represented in the null.

\subsection{Other methods}

The FTP is not the only technique sensitive to non-sinusoidal periodicity. The autocorrelation function makes no assumption about the signal shape and measures cycle-to-cycle memory \citep[e.g.,][]{edelson1988}. Gaussian-process regression with quasi-periodic kernels accommodates stochastic deviations from strict periodicity \citep{foremanmackey2017, covino2020}, and wavelet analyses track the lifetime and frequency drift of a signal \citep{foster1996, torrence1998}. These approaches quantify the persistence of a periodicity, whereas the FTP evaluates its intra-cycle morphology, and the two are complementary. First-lag autocorrelation statistics are already included as out-of-band features in our classifier (Sect.~\ref{subsec:ml_setup}), through which shape-independent coherence information contributes to the ranking.

\section{Possible origins of burst-dominated periodic morphology}
\label{sec:physics}

We make no periodicity claim, and the following remarks are therefore conditional. They concern light curves in which the emission is concentrated in a small fraction of each cycle, the shape that a narrow burst template is designed to match.

Such profiles can arise geometrically. For a relativistically beamed emission region moving along a helical path, the observed flux scales as a high power of the Doppler factor. A narrow viewing angle then turns a smooth rotation into a brief spike each time the velocity vector crosses the line of sight \citep{camenzind1992, rieger2004}. A related possibility is a structured jet in which periodic changes in viewing angle alternately amplify a slow sheath and expose a faster, more variable spine \citep{ghisellini2005, raiteri2017, madero2026, dominguez2026}. In an SMBBH system, a secondary whose orbit is inclined to the accretion disk of the primary crosses the disk twice per orbit, producing short episodes separated by long quiescent intervals \citep{valtonen2008}. Burst-dominated profiles do not need to be geometric, however. Magnetic reconnection is a threshold process in which energy accumulates slowly and is released abruptly, and a periodic driver would yield the same morphology \citep{giannios2009}. Discriminating among these scenarios requires spectral and multiwavelength information that a $\gamma$-ray timing analysis does not provide.

\section{Summary and conclusions}
\label{sec:conclusions}

We searched 101 seven-day binned \textit{Fermi}-LAT light curves from the LCR with the FTP, using three templates (sinusoidal, narrow Gaussian burst, and empirical asymmetric), and ranked the sources with an ML classifier. We conclude the following.

(1) With the quality cuts, three sources exceed a per-source significance of $3\sigma$, against $0.14$ expected for the sample, and one of them is PG 1553+113. No source reaches $3\sigma$ after the correction for the number of light curves searched, and we report no new periodicity detection. The 14 sources with a local significance of at least $3\sigma$ in the first pass match the 12.1 expected from the same rule applied to the synthetic light curves.

(2) The template changes the recovered significance substantially. For 4FGL J0102.8+5824 the per-source significance is $0.24\sigma$ with a sinusoid and $3.85\sigma$ with a Gaussian burst. Differences of this size are expected for about one of the 101 sources in simulated red noise, and $\Delta\sigma$ alone does not establish a periodic signal. The observed distribution of $\Delta\sigma$ has a heavier tail than the null, consistent with light curves that are more burst-dominated than Gaussian red noise, and the single-template FAPs show the same excess at small values. About 40\% of the sources are best described by the sinusoid. Non-sinusoidal templates are therefore a useful complement to the sinusoid in periodicity searches, not a replacement for it.

(3) The classifier, which combines periodogram features with measures of phase coherence and period stability, recognizes periodic signals in simulations but also responds to isolated flares: the number of observed sources scoring above $0.5$ is consistent with that of simulated red noise with non-recurring flares and far above that of Gaussian red noise. It does not separate periodic structure from flares in these data, and it provides a third indication that the light curves contain flares absent from the null. A classifier of this kind needs a noise class as realistic as the significance calibration. Its score is a ranking statistic and not a probability.

(4) Our results agree with recent systematic searches, which find no statistically significant population of periodic $\gamma$-ray blazars. For the nine sources with a local significance of at least $3\sigma$ that also appear in \citet{rico2025} or \citet{penil2026}, the period we recover agrees with a published value. This is not independent evidence, since the analyses use largely the same photons.

(5) Periodicity searches on LCR light curves depend on the convergence criterion of the LCR documentation, which a simple consistency check on $F/\sqrt{\text{TS}}$ complements. A few bins with unconstrained likelihood fits flatten the periodogram of a light curve and leave its fitted spectral index unconstrained or near zero, which makes the simulated null effectively white. The cuts of Sect.~\ref{subsec:sample} remove them, and the dynamic range of the periodogram and the number of threshold crossings reveal the problem at negligible cost. These effects are larger for 7-day than for 30-day binning (Sect.~\ref{subsec:dq}).

Future automated periodicity pipelines applied to high-cadence facilities, such as the Vera C. Rubin Observatory and the Cherenkov Telescope Array Observatory (CTAO), will benefit from moving beyond the sinusoidal assumption to uncover the full diversity of periodic phenomena in the time-domain universe.

\begin{acknowledgements}
The authors thank Daniel Nieto Casta\~no for fruitful discussions. AD acknowledges support from MCIN/AEI under grant PID2025-171816NB-I00; A. Dinesh under grant PID2022-138132NB-C42 (MCIN/AEI/10.13039/501100011033/FEDER, EU).

The $Fermi$-LAT Collaboration acknowledges generous ongoing support from a number of agencies and institutes that have supported both the development and the operation of the LAT as well as the scientific data analysis. These include the National Aeronautics and Space Administration and the Department of Energy in the United States, the Commissariat à l'Energie Atomique and the Centre National de la Recherche Scientifique / Institut National de Physique Nucléaire et de Physique des Particules in France, the Agenzia Spaziale Italiana and the Istituto Nazionale di Fisica Nucleare in Italy, the Ministry of Education, Culture, Sports, Science and Technology (MEXT), High Energy Accelerator Research Organization (KEK) and Japan Aerospace Exploration Agency (JAXA) in Japan, and the K. A. Wallenberg Foundation, the Swedish Research Council and the Swedish National Space Board in Sweden. Additional support for science analysis during the operations phase is gratefully acknowledged from the Istituto Nazionale di Astrofisica in Italy and the Centre National d'Études Spatiales in France. This work performed in part under DOE Contract DE-AC02-76SF00515. This research has made use of the LCR, hosted at the Fermi Science Support Center. We acknowledge the work of the LCR developers and maintainers; specifically, we reference Abdollahi et al. (2023) for the use of results presented in the repository.
\end{acknowledgements}


\section*{Data availability}
Data are available in the \textit{Fermi}-LAT Light Curve Repository.

\appendix
\nolinenumbers

\section{Sources with a local significance of at least $3\sigma$}
\label{app:local}

Table~\ref{tab:local} lists the sources with a local significance of at least $3\sigma$ in the first pass. There are 14, against 12.1 expected when the same rule is applied to the synthetic light curves of the calibration, with a probability of 0.32 of finding at least as many. The number of local $3\sigma$ crossings is therefore what the null predicts, and the sources in the table are not individually evidence of periodicity. The significance of 4FGL J0501.2$-$0158 should be read with care, because its periodogram is nearly flat (dynamic range $1.9$) and the spectral index of its null is poorly constrained. The table also gives the periods and significances published by \citet{rico2025} and \citet{penil2026} for the nine sources that appear in those works (Sect.~\ref{subsec:lit}).

\input{appendix_local}

\end{document}

%% file: table1.tex
\begin{table*}
\caption{Sources within $0.3\sigma$ of the $2.97\sigma$ ceiling of the first pass ($2\times10^{3}$ realizations), re-evaluated with $10^{5}$ realizations. $\sigma_{\rm 2k}$ is the highest first-pass per-source significance. The following columns refer to the $10^{5}$ realizations: $\sigma_{\rm Sine}$, $\sigma_{\rm Gauss}$, and $\sigma_{\rm Emp}$ are the per-source significances for each template, $\sigma_{\rm best}$ is the highest of the three, and $\Delta\sigma = \sigma_{\rm best} - \sigma_{\rm Sine}$. $\sigma_{\rm glob}$ is the global significance, which includes the trial factor for the 101 light curves searched. $\sigma_{\rm loc}$ is the local significance, with no trial factor over periods or templates, and the symbol $>$ marks lower limits at the ceiling of the simulations ($4.26\sigma$). $S_{\rm morph}$ is the classifier ranking score and not a probability.}
\label{tab:results}
\centering
\setlength{\tabcolsep}{4pt}
\begin{tabular}{llccccccccccc}
\hline\hline
4FGL Name & Association & $\alpha$ & $\sigma_{\rm loc}$ & $\sigma_{\rm 2k}$ & $\sigma_{\rm Sine}$ & $\sigma_{\rm Gauss}$ &
$\sigma_{\rm Emp}$ & $\sigma_{\rm best}$ & Template & $\Delta\sigma$ &
$\sigma_{\rm glob}$ & $S_{\rm morph}$ (\%) \\
\hline
J0102.8+5824 & TXS 0059+581 & 0.83 $\pm$ 0.19 & $>$4.26 & 2.97 & 0.24 & 3.85 & 1.71 & 3.85 & Gauss & +3.61 & 2.51 & 81 \\
J1555.7+1111 & PG 1553+113 & 0.54 $\pm$ 0.12 & $>$4.26 & 2.97 & 3.85 & 3.62 & 3.85 & 3.85 & Sine & +0.00 & 2.51 & 79 \\
J0811.4+0146 & OJ 014 & 0.53 $\pm$ 0.33 & 4.26 & 2.97 & 3.34 & 2.06 & 3.38 & 3.38 & Emp & +0.04 & 1.80 & 11 \\
\hline
\end{tabular}
\end{table*}

%% file: appendix_local.tex
\begin{table*}
\caption{Sources with a local significance of at least $3\sigma$ in the first pass ($2\times10^{3}$ realizations, local ceiling $3.29\sigma$; the symbol $>$ marks lower limits at the ceiling). For the three sources of Table~\ref{tab:results} the values refer to the $10^{5}$ realizations (ceiling $4.26\sigma$). $\sigma_{\rm loc}$ is the highest local significance of the three templates, $P$ is the period of the corresponding peak, $\sigma_{\rm src}$ is the per-source significance, $\sigma_{\rm glob}$ is the global significance (values below $1\sigma$ are shown as $<1$, as in \citealt{rico2025}), and $S_{\rm morph}$ is the classifier ranking score and not a probability. The last two columns give, for the sources that also appear in \citet{rico2025} or \citet{penil2026}, the published period in years followed by the significances as published: for \citet{rico2025}, the local significance, which includes the search over periods, and the global significance for 494 sources; for \citet{penil2026}, the median local significance over methods and noise models. Two periods joined by a dash are two periods listed in the original table. These significances are not equivalent to ours or to each other. For S3 0458$-$02, the listed period is the most significant one of \citet{rico2025}; they also report a period of $3.8$~yr with a significance below $2\sigma$, which is not used in the comparison. $^{a}$\citet{penil2025c} report a period of $4.0$~yr for this source (Sect.~\ref{subsec:lit}).}
\label{tab:local}
\centering
\footnotesize
\setlength{\tabcolsep}{3.5pt}
\begin{tabular}{llcccccccc}
\hline\hline
\noalign{\smallskip}
4FGL Name & Association & $\sigma_{\rm loc}$ & Template & $P$ (yr) & $\sigma_{\rm src}$ & $\sigma_{\rm glob}$ & $S_{\rm morph}$ (\%) & Rico et al. (2025) & Pe\~nil et al. (2026a) \\
\noalign{\smallskip}
\hline
\noalign{\smallskip}
J0102.8+5824$^{a}$ & TXS 0059+581 & $>$4.26 & Gauss & 4.36 & 3.85 & 2.51 & 81 & $-$ & $-$ \\
J0211.2+1051 & MG1 J021114+1051 & $>$3.29 & Gauss & 2.84 & 1.96 & $<1$ & 40 & $-$ & $-$ \\
J0501.2$-$0158 & S3 0458$-$02 & $>$3.29 & Sine & 3.97 & 1.98 & $<1$ & 22 & 1.00 (4.8; 3.3) & 4.0 (2.0) \\
J0811.4+0146 & OJ 014 & 4.26 & Sine & 4.31 & 3.38 & 1.80 & 11 & 4.65 (4.8; 3.3) & 4.2 (2.5) \\
J1048.4+7143 & S5 1044+71 & $>$3.29 & Sine & 3.17 & 2.61 & $<1$ & 99 & 3.05 (4.8; 3.3) & 3.1 (3.7) \\
J1058.4+0133 & 4C +01.28 & $>$3.29 & Gauss & 3.71 & 2.37 & $<1$ & 5 & $-$ & $-$ \\
J1146.9+3958 & S4 1144+40 & $>$3.29 & Gauss & 3.12 & 1.56 & $<1$ & 98 & 3.31 (4.8; 3.3) & 3.1 (2.2) \\
J1253.2+5301 & S4 1250+53 & $>$3.29 & Sine & 2.20 & 1.79 & $<1$ & 1 & 2.31 (2.6; $<$1.0) & 2.2 (1.5) \\
J1427.0+2348 & PKS 1424+240 & $>$3.29 & Sine & 1.48 & 0.50 & $<1$ & 6 & 1.53 (4.5; 3.0) & $-$ \\
J1555.7+1111 & PG 1553+113 & $>$4.26 & Sine & 2.09 & 3.85 & 2.51 & 79 & 2.15 (4.8; 3.3) & 2.1 (3.7) \\
J1800.6+7828 & S5 1803+784 & 3.09 & Sine & 1.13 & 0.09 & $<1$ & 2 & $-$ & $-$ \\
J1806.8+6949 & 3C 371 & $>$3.29 & Sine & 2.19 & 2.31 & $<1$ & 96 & $-$ & $-$ \\
J2158.8$-$3013 & PKS 2155$-$304 & 3.09 & Sine & 1.71 & 0.43 & $<1$ & 44 & 1.65 (4.5; 3.0) & 1.7 (1.7) \\
J2236.3+2828 & B2 2234+28A & 3.29 & Gauss & 2.57 & 1.53 & $<1$ & 16 & 2.19 (4.1; 2.3) & 1.4--2.9 (0.5) \\
\noalign{\smallskip}
\hline
\end{tabular}
\end{table*}